# Adaptive Spatiotemporal Dimension Reduction in Concurrent Multiscale Damage Analysis

Shiguang Deng [a], Diran Apelian [a], Ramin Bostanabad [b, *]

[a] ACRC, Materials Science and Engineering, University of California, Irvine, CA, USA
[b] Mechanical and Aerospace Engineering, University of California, Irvine, CA, USA

**Abstract**

Concurrent multiscale damage models are often used to quantify the impacts of manufacturing-induced micro-porosity on the damage response of macroscopic metallic components. However, these models are challenged by major numerical issues including mesh dependency, convergence difficulty, and low accuracy in concentration regions. In this paper, we make two contributions to address these difficulties. Firstly, we develop a novel adaptive assembly-free implicit-explicit (AAF-IE) temporal integration scheme for nonlinear constitutive models. This scheme prevents the convergence issues that implicit algorithms face amid softening. Our AAF-IE scheme autonomously adjusts step sizes to capture intricate history-dependent deformations. It also dispenses with re-assembling the stiffness matrices in elasto-plasticity and damage models which, in turn, dramatically reduces memory footprints. Secondly, we propose an adaptive clustering-based domain decomposition strategy to dramatically reduce the spatial degrees of freedom by agglomerating close-by finite element nodes into a limited number of clusters. Our adaptive clustering scheme has static and dynamic stages that are carried out during offline and online analyses, respectively. The adaptive strategy updates the cluster density based on the spatial discontinuity of the plastic strain. As demonstrated by numerical experiments, the proposed adaptive method strikes a good balance between efficiency and accuracy for fracture simulations. In particular, we use our efficient concurrent multiscale model to quantify the significance of spatially varying microscopic porosity on a macrostructure's softening behavior.

## 1. Introduction

Cast alloys have heterogeneous material properties which are primarily due to manufacturing-induced defects. Process-induced pores are critical defects that generally appear due to gas or shrinkage [1], [2]. These pores are non-uniformly distributed in a metallic component and typically possess complex spatially varying morphologies, see Figure 1(a). Porosity deteriorates a component's structural integrity and load-carrying capacity where alloys are ultimately fractured by crack propagations through pores [3]–[5]. It is, therefore, crucial to quantify the effects of local porosity on a macrostructure's damage response. This quantification is typically achieved via multiscale models that leverage the scale separation between macro-components and micro-pores. While these models are quite powerful, in the presence of softening they become prohibitively expensive, memory demanding, and error prone. In this paper, we aim to address these challenges via an adaptive reduced-order multiscale damage model that predicts the strain-softening behaviors of manufactured alloys with complex local porosity defects.

---

* Corresponding author.
  E-mail address: raminb@uci.edu (R. Bostanabad).

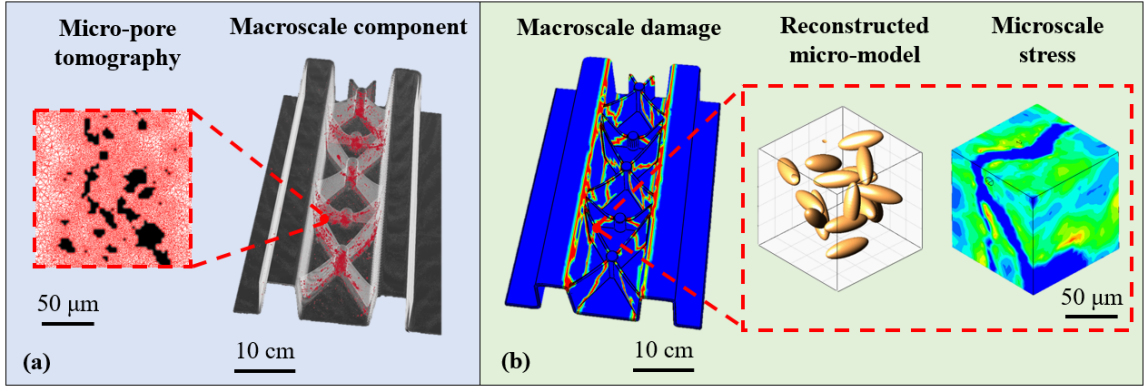

**Figure 1 Multiscale modeling with damage:** (a) A metallic component contains non-uniformly distributed microscale pores whose morphologies are identified via 3D X-ray tomography. (b) The distribution of the damage patterns (fracture bands) on the macro-component is strongly affected by its spatially varying porous microstructures.

*1.1. Background and relevant works*

The influence of micro-porosity on alloys' macroscopic responses can be quantified by various multiscale models [6] such as the homogenization-based methods which leverage the scale-coupling theory. This theory involves the solutions of two nested boundary value problems (BVPs) between macroscopic integration points (IPs) and the associated microstructures: at each increment of the solution process, macro deformation gradients are used to build and solve an admissible microstructural BVP whose homogenized solutions provide the macro variables for the next increment. The coupled iterations continue till both scales achieve equilibriums. The major advantage of multiscale models is the explicit treatment of microstructure which dispenses with the need to devise phenomenological constitutive laws [7].

The concept of representative volume element (RVE) [8] plays an essential role in multiscale modeling as microscale BVPs are solved in such domains. The size of an RVE should be sufficiently large such that a larger domain with different local morphologies provides the same homogenized behavior. In most practical applications, however, the large RVE size and the embedded morphological details result in high computational costs, especially for the finite element method (FEM) [9]. An efficient alternative to FEM is the method of fast Fourier transformation (FFT) [10] which typically uses a homogeneous auxiliary elastic material as the reference medium to predict the effective responses of heterogeneous microstructures with geometrical or material nonlinearity. FFT relies on voxelated microstructure representation which limits this method's ability to capture fine geometrical details. Additionally, FFT does not apply to microstructures with infinite moduli contrast between material phases since elastic tensors lose coerciveness in such cases and, in turn, result in non-unique full-field solutions [11].

To reduce the simulation costs of FEM and FFT, reduced-order models (ROMs) are developed where the key idea is to systematically reduce the number of unknown variables via offline calculations. Transformed field analysis (TFA) [12] and its later version non-uniform transformed field analysis (NTFA) [13] are two of the earliest ROMs. In these two methods, the number of state variables is reduced by expressing arbitrary strain fields as a function of precomputed eigenstrains.

Clustering-based ROMs are recent approaches for predicting the nonlinear responses of heterogeneous materials. Self-consistent analysis (SCA) [14] presumes that elements with similar elastic responses undergo similar plastic deformations. These clusters are generated by grouping voxels with similar elastic responses and the cluster-to-cluster interaction is quantified by incremental Lippmann-Schwinger (LS) equations. The number of unknown variables in SCA is significantly smaller than that in FEM since it assumes identical states per cluster. Virtual clustering analysis (VCA) [15], a variant of SCA, further improves efficiency by avoiding outer loop iterations and considering boundary terms ignored in LS equations. FEM-cluster-based analysis (FCA) [16] constructs the interaction matrix based on clusters' eigenstrains offline and approximates microstructural effective properties online by following the principle of cluster minimum complementary energy. In deflated clustering analysis (DCA) [17], a clustering-based domain decomposition is universally applied to macro- and micro- domains to accelerate, respectively, high-fidelity calculation of macroscale deformations and effective microscopic responses. Its clusters are created by agglomerating close-by nodes where high-fidelity macroscale solutions are accelerated via deflated methods and effective microstructural responses are expedited in a coarse-graining manner such that nodal states in each cluster take on the same value.

Although ROMs are quite efficient in simulating elasto-plastic deformations, their successful application in modeling fracture is rare due to the intricate deformation mechanics. Fracture is generally modeled by two different approaches: (1) a fracture mechanics-based discrete approach that directly simulates displacement discontinuity between fracture interfaces using elastic mediums [18], and (2) a continuum mechanics approach that emulates fractures via strain softening using localized plastic strains. While most ROMs are developed in the realm of continuum mechanics, their applications in the second fracture simulation approach are obstructed by softening-induced numerical instabilities. Specifically, the stiffness matrices of damaged materials lose positive-definiteness amid crack propagation causing ill-conditioned equilibrium equations with imaginary wave speeds [19]. If not properly resolved, solutions of the ill-posed fracture model exhibit spurious mesh dependency that restricts damage elements to single-element wide fracture bands and causes unphysically diminishing energy dissipations upon mesh refinements. Numerical remedies include crack band theory [20], non-local models [21], and phase-field methods [22]. These remedies either introduce mesh-dependent regularizations or modify damage evolution to spread the fracture over finite regions.

Another difficulty in simulating softening via ROMs is the convergence of pure implicit time integration schemes. In these schemes, convergence issues arise due to the non-positive-definite stiffness matrices which cause near-zero or negative gradients in incremental algebraic systems. One way to address this issue is to use explicit time integration schemes which integrate the equations of motion explicitly through time where current kinematic solutions are directly extrapolated from previous steps. However, explicit schemes are very expensive since they require much smaller time steps compared to implicit schemes.

To improve the emulation accuracy of strain-softening, adaptive discretization is often integrated with damage models to enrich spatial interpolations at localized regions while maintaining manageable computational costs. A typical example is adaptive meshing in FEM which is classified as either h-adaptivity or p-adaptivity [23] (in h-adaption element types are unchanged while p-adaption adjusts mesh types by assigning high-order interpolations in critical regions and coarse interpolations elsewhere). Adaptive meshing is also utilized in multiscale damage models [24]–[26] which often involves identifying the critical regions in the macro-domain where concentrations are expected to develop. The difficulties associated with such an approach include precise anticipation of the critical regions, efficient swap between coarse and fine meshes, and coupling non-matching meshes on the boundaries of subdomains to enforce displacement compatibility. These difficulties can be alleviated by using periodic unit cells with congruent boundaries over different subdomains and enforcing displacement compatibility by Lagrangian multipliers [27]. The adaption strategy is only recently integrated with ROMs [28] where spatial refinements are adaptively performed over material clusters during plastic deformations. These refinements rely on an iterative procedure that, based on a heuristic error metric, rewinds the simulation state to a previous time step where state variables are recalculated on a finer discretization.

*1.2. Research gaps and our contributions*

We summarize the research gaps in damage modeling for multiscale metallic components as follows:
- Existing methods primarily rely on direct numerical simulations (DNS) to emulate macroscopic damage behaviors with associated microscale heterogeneities. Since these methods are generally memory demanding and computationally expensive, efficient ROMs with high accuracy are needed to quantify the effect of micro pores on the macro component response.
- Most existing ROMs mainly focus on nonlinear elasto-plastic deformations that involve strain-hardening and their successful applications for modeling damage simulations are limited due to softening-induced numerical instabilities. New numerical methods are needed to enable ROMs to resolve softening efficiently and robustly.
- While clustering-based ROMs are highly efficient in approximating microstructures' effective responses, they lose accuracy in concentration regions with highly localized softening. Advancements in ROMs are needed to achieve global and local accuracy.

To fill these gaps, we propose a novel ROM to simulate the damage behavior of metallic alloys with process-induced microscopic porosity. Our method is called adaptive deflated clustering analysis (ADCA) and involves two major contributions. Firstly, it involves a new adaptive assembly-free impl-exp (AAF-IE) temporal integration scheme which resolves the softening-induced convergence issues by preserving the positive-definiteness of the underlying algebraic system amid damage evolution. Our AAF-IE does not require online re-assembly of the global stiffness matrix which dramatically improves computational efficiency. Secondly, we develop a novel adaptive spatial discretization strategy which consists of a static offline clustering stage and an online dynamic stage. While the offline stage assigns high cluster densities in critical regions identified through inexpensive elastic loadings, the online procedure adaptively adjusts cluster densities over crucial areas where softening is expected to initiate and propagate. We integrate the online adaption stage with a novel error metric to automatically detect spatial regions and temporal instances (i.e., there is no need for rewinding the simulation back to an earlier instance) that need finer interpolations. To prevent softening-induced spurious mesh dependencies on multiple scales, we regularize damage evolutions with macroscopic non-local functions and fracture energy stabilized micro-damage models, both of which are systematically integrated with our AAF-IE. It is noted that while we integrate our adaptive techniques with DCA [17], they can be readily integrated with any other clustering-based ROM for efficient multiscale modeling with accurate predictions of localized phenomena.

*1.3. Outline of the paper*

The remainder of the paper is organized as follows. In Section 2, we briefly review some background techniques. Our adaptive multiscale damage method is introduced in Section 3 and we demonstrate its efficiency and accuracy via numerical experiments in Section 3.3. The paper is concluded with some final remarks in Section 5.

## 2. Background techniques

The proposed reduced-order multiscale damage model relies on a few background techniques including homogenization theory, damage models, and time integration schemes which are briefly reviewed below.

*2.1. Multiscale modeling via first-order homogenization*

We use multiscale models to investigate the influence of microscale heterogeneity on the fracture behavior of macroscale components. Our multiscale model relies on the first-order computational homogenization (FOCH) method which assumes distinguishable macro and micro features. Specifically, the sizes of local material heterogeneity ($l_\mu$) is assumed to be much smaller than the RVE size ($l_m$) which itself is presumed to be smaller than the macro-characteristic length ($l_M$), that is:

$$l_\mu \ll l_m \ll l_M \tag{1}$$

where the subscripts 'M' and 'm' represent macroscale and microscale, respectively.

Solutions of the macro- and micro-structures are coupled via the Hill-Mandel energy condition [29] which implies that the virtual internal work density at a macro IP equals the volume average of the virtual work of its associated RVE:

$$\mathbf{S}_M : \delta \mathbf{E}_M = \frac{1}{|\Omega|} \int_\Omega \mathbf{S}_m : \delta \mathbf{E}_m d\Omega \tag{2}$$

where $\mathbf{S}_M, \mathbf{E}_M, \mathbf{S}_m$ and $\mathbf{E}_m$ are the macroscopic and microscopic stress and strain tensors, respectively. $\Omega$ and $|\Omega|$ represent the domain of microstructure and its volume. The symbol ':' represents the double dot product that contracts a pair of repeated indices.

The stress and strain tensors at either scale are computed through equilibrium equations. The equilibrium equations for a macrostructure in the infinitesimal deformation framework are [30]:

$$\mathbf{S}_M(\mathbf{X}) \cdot \nabla_0 + \mathbf{b}_M = \mathbf{0} \qquad \forall \mathbf{X} \in \Omega_{0M} \tag{3}$$

$$\mathbf{u}_M(\mathbf{X}) = \bar{\mathbf{u}}_M \qquad \forall \mathbf{X} \in \Gamma_{0M}^D \tag{4}$$

$$\mathbf{S}_M(\mathbf{X}) \cdot \mathbf{n}_M = \bar{\mathbf{t}}_M \qquad \forall \mathbf{X} \in \Gamma_{0M}^N \tag{5}$$

where $\mathbf{u}_M$ is the unknown macro-displacement, $\bar{\mathbf{u}}_M$ is the prescribed displacement on the Dirichlet boundary $\Gamma_{0M}^D$, $\bar{\mathbf{t}}_M$ represents the surface traction over the Neumann boundary $\Gamma_{0M}^N$, and $\mathbf{n}_M$ denotes the outward unit vector to the boundary of the undeformed macro-domain $\Omega_{0M}$. $\nabla_0$ is the gradient operator with respect to the undeformed configuration.

Similarly, the equilibrium equation in the microscale can be written as the following BVP:

$$\mathbf{S}_m(\mathbf{x}) \cdot \nabla_0 = \mathbf{0} \qquad \forall \mathbf{x} \in \Omega_{0m} \tag{6}$$

$$\mathbf{S}_m(\mathbf{x}) \cdot \mathbf{n}_m = \bar{\mathbf{t}}_m \qquad \forall \mathbf{x} \in \Gamma_{0m}^N \tag{7}$$

where $\bar{\mathbf{t}}_m$ is the prescribed surface traction per unit area on the boundary $\Gamma_{0m}^N$ of reference microscale domain with the outward unit normal vector $\mathbf{n}_m$. According to the Hill-Mandel condition, macroscale and microscale equations are coupled by equating the macroscopic stress to the homogenized micro-stress of RVE and storing history-dependent state variables for continued analyses in the concurrent multiscale model.

We assume the plastic behavior of the studied alloy system follows a rate-independent isotropic elasto-plastic constitutive law in the microscale as:

$$\begin{cases} \mathbf{S}_m = \mathbb{C}^{el} : \mathbf{E}_m^{el} \\ \mathbf{E}_m^{el} = \mathbf{E}_m - \mathbf{E}_m^{pl} \\ \mathbf{E}_m = \int d\mathbf{E}_m \\ \mathbf{E}_m^{pl} = \int d\mathbf{E}_m^{pl} \end{cases} \tag{8}$$

where $\mathbb{C}^{el}$ represents the fourth-order elasticity tensor. The total microscopic strain ($\mathbf{E}_m$) is additively decomposed into the microscopic elastic strain ($\mathbf{E}_m^{el}$) and plastic strain ($\mathbf{E}_m^{pl}$) where the values of the total and plastic strains are integrated by their increments ($d\mathbf{E}_m$ and $d\mathbf{E}_m^{pl}$) at each time step. The plastic state is defined by the yield condition $f$ as:

$$f = f(\mathbf{S}_m, q) = 0 \tag{9}$$

where $q$ is a history-dependent state variable. During plastic flow, the plastic strain increment $\Delta \mathbf{E}_m^{pl}$ is determined by the plastic flow rule as:

$$\Delta \mathbf{E}_m^{pl} = \Delta \lambda \frac{\partial Q}{\partial \mathbf{S}_m} \tag{10}$$

where $\lambda$ is the plastic multiplier and $Q$ refers to the plastic potential. In this work, we use an associated flow rule where $Q = f$.

## 2.2. Damage in ductile metals

Damage occurs when excessively large loads are applied to materials whose loading-carrying capacity is reduced as a result of progressive degradation of the materials' moduli. To model this degradation in a multiscale setting, we add damage to Equations (3)-(7). In particular, we adopt an isotropic continuum damage model which has two major components: an initiation criterion that predicts the onset of softening and an evolution law that traces the crack's progression until rupture.

We select the ductile criterion at either scale as the damage initiation metric which assumes the effective plastic strain ($\bar{E}_d^{pl}$) at the onset of fracture is a function of stress states and strain rates. For simplicity, we assume $\bar{E}_d^{pl}$ is a user-defined constant and damage is initiated when the following condition is met:

$$\omega_d = \frac{\overline{E}^{pl}}{\overline{E}_d^{pl}} \geq 1 \tag{11}$$

where $\omega_d$ is the damage initiation variable that increases monotonically in elasto-plastic deformations and $\overline{E}^{pl}$ is the equivalent plastic strain. We note that ductile damage initiation can be used in conjunction with other damage initiation criteria. When multiple criteria are specified simultaneously, they are treated independently, and damage evolution starts if any criterion is met.

Upon initiation, damage results in softening of the yield stress and degradation of the stiffness. Correspondingly, the constitutive equation for a damaged elasto-plastic metal is written as:

$$\begin{cases} \mathbf{S} = (1-D)\mathbf{S}^0 \\ \mathbf{S}^0 = \mathbb{C}^{el} : \mathbf{E}^{el} = \mathbb{C}^{el} : (\mathbf{E} - \mathbf{E}^{pl}) \end{cases} \tag{12}$$

where $D \in [0, 1]$ is the monotonically increasing damage evolution parameter and $\mathbf{S}^0$ is the stress tensor on a reference material that undergoes the same plastic deformation but in the absence of damage. Since the continuum damage is assumed as isotropic, $D$ is a scalar in Equation (12). If multiple evolution laws are specified, $D$ would capture the effects of all active damage mechanisms. For an anisotropic damage model, however, $D$ becomes a tensor.

The evolution of $D$ can be specified as a function of equivalent plastic strain and state variables, e.g., as an exponential function [31] of $\overline{E}^{pl}$ and a user-defined non-negative evolution rate parameter ($\alpha$) as:

$$D(\overline{E}^{pl}, \alpha) = 1 - \frac{\overline{E}_d^{pl}}{\overline{E}^{pl}} \exp(-\alpha(\overline{E}^{pl} - \overline{E}_d^{pl})) \tag{13}$$

This definition ensures that when $\overline{E}^{pl}$ is sufficiently large compared to $\overline{E}_d^{pl}$, the damage variable increases to one so that the stiffness is fully degraded, and the material is entirely ruptured.

Once the maximum degradation is reached at a material point, a choice of removal of the ruptured elements from the computational mesh is enabled. Element deletion prevents further damage accumulation in localized regions and enhances computational efficiency by improving the condition number of damaged stiffness of the underlying algebraic systems.

*2.3. Mesh dependency control*

With fractures formulated as strain softening by the isotropic continuum damage model in Equation (12), softening-induced non-positive definite stiffness causes the BVPs in Equation (3)-(7) to be ill-conditioned with unstable convergence and negative wave speeds. In such a scenario, equilibrium solutions lose their objectivity with respect to mesh sizes and exhibit spurious mesh sensitivity, especially when using standard finite elements (FEs) of the first-order continuity. If not properly addressed, the progressive damage model in Equation (12) would result in unphysical material failures that are localized in a certain region with shrinking fracture energies upon mesh refinements [21]. As discussed below, we adopt two different approaches to control the mesh dependency on the macro and micro domains.

*2.3.1. Macroscale mesh dependency control by nonlocal functions*

When the damage model of Equation (12) is directly applied to the damage evolutions in Equation (13), macrostructures exhibit unphysical material softening in concentration regions where the equilibrium solutions strongly depend on the mesh. This pathological mesh dependency restricts the fractured elements into an unphysical narrow damage band with a single layer of elements. This spurious mesh dependency can be alleviated by introducing the 'fracture band width' as a macroscale material characteristic length in a nonlocal damage model.

We adopt an integral-type nonlocal damage model in Equation (14) where the nonlocal parameter of a macroscopic point involves the weighted averages of the damage parameters over a finite spatial neighborhood of the point under consideration:

$$\hat{D}(\mathbf{X}, \mathbf{X}') = \int_B \omega(\|\mathbf{X} - \mathbf{X}'\|) D(\mathbf{X}') d(\mathbf{X}') \tag{14}$$

where $\hat{D}(\mathbf{X}, \mathbf{X}')$ is the nonlocal damage parameter at a macroscopic point $\mathbf{X}$ surrounded by neighbor points $\mathbf{X}'$ in a support neighborhood denoted by B, $D(\mathbf{X}')$ is the local damage parameter at $\mathbf{X}'$, and $\omega(\|\mathbf{X} - \mathbf{X}'\|)$ is the nonlocal weighting function. There is no unique way to define $\omega(\|\mathbf{X} - \mathbf{X}'\|)$ and in this study, we adopt a polynomial bell-shaped function [31] as:

$$\omega(\|\mathbf{X} - \mathbf{X}'\|) = \frac{\omega_\infty(\|\mathbf{X} - \mathbf{X}'\|)}{\int_B \omega_\infty(\|\mathbf{X} - \mathbf{X}'\|) d(\mathbf{X}')} \tag{15}$$

$$\omega_\infty(\|\mathbf{X} - \mathbf{X}'\|) = \left\langle 1 - \frac{4(\|\mathbf{X} - \mathbf{X}'\|)^2}{l_0^2} \right\rangle^2 \tag{16}$$

where the Macauley brackets $\langle ... \rangle$ represent a non-negative value defined as $\langle x \rangle = \max(0, x)$, and $l_0$ is the macroscale material characteristic parameter denoting an interaction radius, that is, the weighting effects diminish when $\|\mathbf{X} - \mathbf{X}'\| > l_0/2$. Thus, in 3D models, the support domain B is a sphere with a radius of $l_0/2$.

The macroscale material characteristic length $l_0$ determines the width of fracture bands whose value can be measured by high-fidelity numerical simulations using discrete fracture mechanics or via a dedicated experiment with high-resolution digit image correlation analyses. To check mesh independence in fracture analyses, element sizes are generally selected to be much smaller than $l_0$ and the post-peak damage responses are tracked to show convergence upon mesh refinement, which we demonstrate in Sections 4.1 and 4.2.

*2.3.2. Microscale mesh dependency control by fracture energy*

In the presence of fracture, microstructure responses are no longer properly represented by stress-strain relations since they cause damage models to lack objectivity to mesh choice and result in imaginary wave speeds amid damage progression. Using an arbitrary microscale characteristic length ($l_1$) in a microscale damage model, similar to the macroscopic counterpart ($l_0$) defined in Equation (16), helps to stabilize ill-posed strain-softening but the simultaneous applications of $l_0$ and $l_1$ in a multiscale model is not physically realistic.

To mitigate mesh dependency on microscopic damage models, the softening part of the constitutive law is converted from the stress-strain relation to a stress-displacement relation by an element characteristic length ($l_e$). Then, the fracture energy ($G_f$) is specified as the dissipated energy (after damage initiation) that opens a unit area of the crack as:

$$G_f = \int_{\bar{E}_0^{pl}}^{\bar{E}_f^{pl}} l_e S_y d\bar{E}^{pl} = \int_0^{\bar{u}_f^{pl}} S_y d\bar{u}^{pl} \tag{17}$$

where $G_f$ is defined on each microscale IP with an $l_e$, and $S_y$ is the yield stress corresponding to $\bar{E}^{pl}$. The equivalent plastic displacement $\bar{u}^{pl}$ is the fracture work conjugate of $S_y$ in the fracture evolution from damage initiation (marked with the effective plastic strain $\bar{E}_0^{pl}$ and zero plastic displacement) to the final rupture (represented by the effective plastic strain $\bar{E}_f^{pl}$ and the fracture plastic displacement $\bar{u}_f^{pl}$). For FEs with the first-order continuity, $l_e$ equals the length of a line across elements. $l_e$ can also be directly specified as a function of element topology or material orientation [32].

With the effective plastic strains converted to plastic displacements with associated elemental $l_e$, the damage evolution is therefore defined based on the released energy during the damage propagation in an exponential form [32] of the plastic displacement as:

$$D = 1 - \exp\left(-\frac{1}{G_f}\int_0^{\bar{u}^{pl}} S_y d\bar{u}^{pl}\right) \tag{18}$$

where the damage variable approaches one only asymptotically at infinitely large plastic displacement. In practice, we set $D$ as one when dissipated energies exceed $0.99 G_f$. We note that fracture directions are generally not known a priori and, as a result, the elements with high aspect ratios behave differently depending on the direction along which the crack occurs. Thus, continuum damage models typically prefer elements with an aspect ratio close to one.

*2.4. Convergence difficulty of implicit solvers*

Strain softening and stiffness degradation amid damage evolution cause serious convergence difficulties for implicit time integration schemes because the implicitly integrated algebraic systems exhibit singular or ill-conditioned stiffness matrices for materials with degraded moduli. To demonstrate this, consider a simple isotropic damage model whose constitutive equation can be integrated analytically [33] by the classic implicit backward-Euler integration scheme. The algorithmic tangent operator $\mathbb{C}_{n+1}^{alg.}$ in the damage model at time step $(n+1)$ is written as [34]:

$$\mathbb{C}_{n+1}^{alg.} \equiv \frac{\partial \mathbf{S}_{n+1}}{\partial \mathbf{E}_{n+1}} = (1 - D_{n+1})\mathbb{C}^{el} - \frac{\mathbf{S}_{n+1} - \mathbf{H}_n \bar{\mathbf{E}}_{n+1}^{pl}}{(\bar{\mathbf{E}}_{n+1}^{pl})^3}\mathbf{S}_{n+1}^0 \otimes \mathbf{S}_{n+1}^0 \tag{19}$$

where $\bar{E}_{n+1}^{pl}$, $S_{n+1}$, $\mathbf{S}_{n+1}^0$, $H_n$ are the equivalent plastic strain, equivalent stress, referenced (undamaged) stress tensor, and softening modulus, respectively (the subscripts denote the time step). Upon damage initiation, $H_n$ becomes negative and hence $\mathbb{C}_{n+1}^{alg.}$ loses its positive definiteness in some loading states. The non-positive definite $\mathbb{C}_{n+1}^{alg.}$ causes the elemental stiffness matrix ($\mathbf{k}_{n+1}^e$) to be ill-conditioned with possible negative eigenvalues. As damaged elements agglomerate in certain strain-softening regions, negative eigenvalues enter the global stiffness matrix ($\mathbf{K}_{n+1}$) via element assembly process:

$$\mathbf{K}_{n+1} = \frac{\partial \mathbf{F}_{n+1}^{int}}{\partial \mathbf{u}_{n+1}} = A_e(\mathbf{k}_{n+1}^e) = A_e\left(\int_{\Omega_e} \mathbf{B}^T : \mathbb{C}_{n+1}^{alg.} : \mathbf{B} d\Omega\right) \tag{20}$$

where $\mathbf{F}_{n+1}^{int}$ and $\mathbf{u}_{n+1}$ are the internal forces and nodal displacements at time step $(n+1)$. $A_e(...)$ is the assembling operator and $\mathbf{B}$ is the strain-displacement matrix evaluated within each element domain $\Omega_e$. So, the locally damaged elements cause the global stiffness matrix to become ill-conditioned. Note that softening occurs not only in damage but also in plastic models, and in both cases, the condition number of the global stiffness matrix deteriorates.

The ill-conditioning of the global stiffness matrix reduces the efficiency of the Newton-Raphson procedure. Specifically, when the algebraic system becomes singular in a certain step due to degraded moduli, the Newton-Raphson iteration halts before convergence. Solutions for improving implicit solvers' convergence rates are based on, e.g., viscous regularization schemes [32] and continuation methods [35] that render the tangent stiffness to be positive-definite in sufficiently small steps. However, such remedies fail in many scenarios and the implicit schemes face severe convergence difficulty. To fundamentally resolve the softening-induced numerical instability, we develop an adaptive hybrid time integration scheme in the next section.

## 3. Proposed adaptive reduced-order multiscale damage model

Our approach has two novel ingredients: (1) an adaptive assembly-free impl-exp time integration scheme that automatically adjusts temporal step sizes amid nonlinear analyses without re-assembling global stiffness matrices of the underlying algebraic system, and (2) an adaptive clustering strategy which alters local spatial discretization to preserve solution accuracy in concentration regions. These temporal and spatial adaption procedures aim to efficiently improve simulation accuracy and they are discussed in Sections 3.1 and 3.2, respectively. The overall steps of our approach are summarized in Section 3.3.

*3.1. Adaptive temporal reduction*

Time-dependent nonlinear analyses are often integrated via pure implicit schemes, e.g., the Newton-Raphson method, due to their high efficiency and unconditional stability. However, in strain-softening simulations, ill-conditioned algebraic systems with near-zero or negative tangent moduli often prevent iterative solutions from convergence even after many steps. To reduce the overall number of temporal steps, we adopt a proper time integration scheme with robust convergence property and improve it as explained below.

*3.1.1. Standard impl-exp scheme*

The impl-exp scheme [34], [36] is a hybrid method that is robust and efficient in integrating softened constitutive equations. This method avoids the softening-induced numerical convergence issues discussed in Section 2.4 and its algorithm is demonstrated for the elasto-plasticity in Table 1 and the ductile damage model in Table 2.

The basic idea of the impl-exp scheme is to maintain the positive-definiteness of the algorithmic tangent operator by dividing constitutive integration into two stages. In the first stage, the plastic multiplier increment $\Delta \tilde{\lambda}_{n+1}$ is explicitly extrapolated at time step $(n + 1)$ from $\Delta \lambda_n$ to compute an 'explicit' stress $\tilde{\mathbf{S}}_{n+1}$ which balances the equilibrium equation between internal and external forces. In the second stage, the 'implicit' stress $\mathbf{S}_{n+1}$ is obtained in terms of the current strain $\mathbf{E}_{n+1}$ via the standard implicit backward Euler method. Then, the 'implicit' stress $\mathbf{S}_{n+1}$ is used in the next step to compute the trial stress for updating yield function, see Table 1. With impl-exp schemes, both tangent operators $\tilde{\mathbb{C}}_{n+1}^{\text{alg.}}$ are guaranteed to be positive-definite for the elasto-plasticity and damage models in Table 1 and Table 2, respectively.

**Table 1**: **Standard impl-exp scheme:** Positive-definiteness of the algorithmic tangent operator is preserved by dividing constitutive integration into two stages.

| Inputs: $\mathbb{C}^{el}, \Delta \mathbf{E}_{n+1}, \mathbf{S}_n, q, Q, \Delta\lambda_n$ | |
|---|---|
| Trial stress: | $\mathbf{S}_{n+1}^{trial} = \mathbf{S}_n + \mathbb{C}^{el}\Delta\mathbf{E}_{n+1}$ |
| Yield criterion: | $f = f^{trial}(\mathbf{S}_{n+1}^{trial}, q)$ |
| Implicit plastic strain increment: | $\Delta\mathbf{E}_{n+1}^{pl} = \Delta\lambda_{n+1}(\partial Q/\partial \mathbf{S}_{n+1})$ |
| Implicit stress: | $\mathbf{S}_{n+1} = \mathbf{S}_{n+1}^{trial} - \mathbb{C}^{el}:\Delta\mathbf{E}_{n+1}^{pl}$ |
| Explicit plastic strain: | $\Delta\tilde{\lambda}_{n+1} = (\Delta t_{n+1}/\Delta t_n)\Delta\lambda_n;\ \tilde{\mathbf{E}}_{n+1}^{pl} = \mathbf{E}_{n+1}^{pl} + \Delta\tilde{\lambda}_{n+1}(\partial Q/\partial \tilde{\mathbf{S}}_{n+1})$ |
| Explicit stress: | $\tilde{\mathbf{S}}_{n+1} = \mathbf{S}_{n+1}^{trial} - \mathbb{C}^{el}\Delta\tilde{\lambda}_{n+1}(\partial Q/\partial \tilde{\mathbf{S}}_{n+1})$ |
| Algorithmic tangent operator: | $\tilde{\mathbb{C}}_{n+1}^{alg.} = \dfrac{\partial \tilde{\mathbf{S}}_{n+1}}{\partial \mathbf{E}_{n+1}} = \dfrac{\partial(\mathbb{C}^{el}(\mathbf{E}_{n+1} - \mathbf{E}_n^{pl}) - \mathbb{C}^{el}\Delta\tilde{\lambda}_{n+1}\frac{\partial Q(\tilde{\mathbf{S}}_{n+1}(\mathbf{E}_{n+1}))}{\partial \tilde{\mathbf{S}}_{n+1}})}{\partial \mathbf{E}_{n+1}}$ |
| Outputs: $\tilde{\mathbb{C}}_{n+1}^{alg.}, \tilde{\mathbf{S}}_{n+1}$ | |

**Table 2**: **Standard impl-exp scheme with damage:** It is assumed that the material behaves elasto-plastically as described in Table 1.

| Inputs: $\mathbb{C}^{el}, \mathbf{E}_{n+1}^{el}$ | |
|---|---|
| Effective stress: | $\mathbf{S}_{n+1}^0 = \mathbb{C}^{el}:\mathbf{E}_{n+1}$ |
| Internal variable: | $\Delta\tilde{\lambda}_{n+1} = (\Delta t_{n+1}/\Delta t_n)\Delta\lambda_n$ |
| Damage initiation criterion: | $\omega_d = \bar{E}_{n+1}^{pl}/\bar{E}_d^{pl} = (\bar{E}_n^{pl} + \Delta\bar{E}_{n+1}^{pl})/\bar{E}_d^{pl} \geq 1$ |
| Damage parameter: | $\tilde{D}_{n+1} = \tilde{D}_{n+1}(D_n, \Delta\tilde{\lambda}_{n+1})$ |
| Damage stress: | $\tilde{\mathbf{S}}_{n+1} = (1 - \tilde{D}_{n+1})\mathbf{S}_{n+1}^0$ |
| Algorithmic tangent operator: | $\tilde{\mathbb{C}}_{n+1}^{alg.} = \partial\tilde{\mathbf{S}}_{n+1}/\partial\mathbf{E}_{n+1} = (1 - \tilde{D}_{n+1})\mathbb{C}^{el}$ |
| Outputs: $\tilde{\mathbb{C}}_{n+1}^{alg.}, \tilde{\mathbf{S}}_{n+1}$ | |

In an impl-exp scheme, the tangent operator is kept constant in a Newton step (as opposed to a pure implicit scheme,) which dramatically reduces the required iteration numbers within a step. However, since the tangent operator varies between steps, the global stiffness matrix must be re-assembled at the first iteration of each Newton step. This repetitive reassembly results in large memory footprints and slows down the overall process. We address this issue in the next subsection with our new method.

*3.1.2. Adaptive assembly-free impl-exp scheme*

To improve the computational efficiency of the impl-exp scheme introduced in Section 3.1.1, we propose a novel adaptive assembly-free impl-exp (AAF-IE) scheme. Compared to the standard impl-exp, our approach is more efficient because it avoids the re-assembly of the underlying stiffness matrix in runtime and adaptively adjusts temporal steps to improve integration accuracy. The key idea is to estimate plastic strain tensors with explicit extrapolations (instead of expensive implicit estimators) while maintaining equilibrium conditions and satisfying constitutive equations. Specifically, in contrast to the standard impl-exp which updates 'explicit' state variables with respect to a linearly extrapolated plastic multiplier increment $\Delta\tilde{\lambda}_{n+1}$, we set the extrapolated state variables in AAF-IE to the plastic strain increment tensor, $\Delta\tilde{\mathbf{E}}^{pl}_{n+1}$, as:

$$\tilde{\mathbf{S}}_{n+1}(\Delta\tilde{\mathbf{E}}^{pl}_{n+1}) = \tilde{\mathbf{S}}^{trial}_{n+1} - \mathbb{C}^{el} : \Delta\tilde{\mathbf{E}}^{pl}_{n+1} = \mathbb{C}^{el} : (\mathbf{E}_n - \mathbf{E}^{pl}_n + \Delta\mathbf{E}_{n+1} - \Delta\tilde{\mathbf{E}}^{pl}_{n+1}) \quad (21)$$
$$= \mathbb{C}^{el} : \mathbf{E}_{n+1} - \mathbb{C}^{el} : \mathbf{E}^{pl}_n - \mathbb{C}^{el} : \Delta\tilde{\mathbf{E}}^{pl}_{n+1}$$

$$\Delta\tilde{\mathbf{E}}^{pl}_{n+1} = \frac{\Delta t_{n+1}}{\Delta t_n} \Delta\mathbf{E}^{pl}_n \quad (22)$$

where $\Delta\mathbf{E}^{pl}_n$ is the converged implicit plastic strain tensor increment from the previous time step ($n$). The algorithmic tangent operator for the elasto-plasticity model is simplified as:

$$\tilde{\mathbb{C}}^{alg.}_{n+1} = \frac{\partial\tilde{\mathbf{S}}_{n+1}(\Delta\tilde{\mathbf{E}}^{pl}_{n+1})}{\partial\mathbf{E}_{n+1}} = \frac{\partial(\mathbb{C}^{el} : \mathbf{E}_{n+1} - \mathbb{C}^{el} : \mathbf{E}^{pl}_n - \mathbb{C}^{el} : \Delta\tilde{\mathbf{E}}^{pl}_{n+1})}{\partial\mathbf{E}_{n+1}} = \mathbb{C}^{el} \quad (23)$$

The advantage of the Equation (23) comes from the fact that the tangent operator $\tilde{\mathbb{C}}^{alg.}_{n+1}$ is now independent of the plastic potential $Q(\tilde{\mathbf{S}}_{n+1}(\mathbf{E}_{n+1}))$, see the standard impl-exp in Table 1, and it always equals to the elastic tangent modulus $\mathbb{C}^{el}$ during the entire analysis. As a result, the global stiffness matrix ($\mathbf{K}_{n+1}$) of the elasto-plastic system in Equation (20) is only assembled once in the offline stage where its Cholesky decomposition matrices are calculated and then stored for repeated usage in runtime. The constant stiffness matrix avoids re-assembly and therefore significantly reduces memory usage while improving efficiency.

As the material becomes damaged, $\tilde{D}_{n+1}$ and $\tilde{\mathbb{C}}^{alg.}_{n+1}$ still depend on the linearly extrapolated $\Delta\tilde{\lambda}_{n+1}$ which itself is a function of the time step increments ($\Delta t_n$ and $\Delta t_{n+1}$) and the implicitly converged $\Delta\lambda_n$. However, since softening is highly localized in a strain-softening zone with a limited number of fractured elements (compared to the whole mesh), the element stiffness matrices of the damaged elements can be incrementally updated in the global stiffness matrix as:

$$\mathbf{K}_{n+1} = \mathbf{K}_n + \Delta\mathbf{K}^d_{n+1} \quad (24)$$

where $\Delta\mathbf{K}^d_{n+1}$ corresponds to the entries with damaged material properties. This incremental assembly technique avoids overall re-assembly and significantly reduces memory footprints.

A major limitation of the standard impl-exp is the accuracy loss incurred by large time steps [34]. The accuracy loss comes from the mismatch between the linearly extrapolated state variables and their real values, especially upon transitioning between elasticity and plasticity where the state variables (e.g., effective plastic strain increments) change abruptly. Thus, it is recommended to use sufficiently small steps to reduce the extrapolation error (like explicit integration schemes) which increases the computational expenses.

Noting that large time steps can be used at non-critical moments when material properties change smoothly, we introduce an adaptive integrator in the framework of an assembly-free impl-exp scheme to automatically adjust the time steps. At the arbitrary time instance $t_{n+1}$, the explicitly extrapolated increment of the plastic strain tensor can be expressed as its implicit counterpart from the last step:

$$\Delta\tilde{\mathbf{E}}^{pl}_{n+1} = \tilde{\mathbf{E}}^{pl}_{n+1} - \mathbf{E}^{pl}_n = \frac{\Delta t_{n+1}}{\Delta t_n} \Delta\mathbf{E}^{pl}_n \quad (25)$$

The Taylor series expansion of the implicit plastic strain at $t_{n-1}$ is:

$$\begin{aligned}
\mathbf{E}_{n-1}^{pl} &= \mathbf{E}_n^{pl}(t_n - \Delta t_n) \\
&= \mathbf{E}_n^{pl} - \dot{\mathbf{E}}_n^{pl}\Delta t_n + \frac{1}{2}\ddot{\mathbf{E}}_n^{pl}(\Delta t_n)^2 - \frac{1}{6}\dddot{\mathbf{E}}_n^{pl}(\Delta t_n)^3 + o((\Delta t_n)^4)
\end{aligned} \quad (26)$$

where $\dot{\mathbf{E}}_n^{pl}$, $\ddot{\mathbf{E}}_n^{pl}$ and $\dddot{\mathbf{E}}_n^{pl}$ are the first, second, and third-order time derivatives at the time-step $n$. The implicit plastic strain increment is then expressed as:

$$\begin{aligned}
\Delta\mathbf{E}_n^{pl} &= \mathbf{E}_n^{pl} - \mathbf{E}_{n-1}^{pl} \\
&= \dot{\mathbf{E}}_n^{pl}\Delta t_n - \frac{1}{2}\ddot{\mathbf{E}}_n^{pl}(\Delta t_n)^2 + \frac{1}{6}\dddot{\mathbf{E}}_n^{pl}(\Delta t_n)^3 + o((\Delta t_n)^4)
\end{aligned} \quad (27)$$

By substituting Equation (27) to (25) and assuming $\Delta t_{n+1} = \Delta t_n$, we arrive at the expansion of the extrapolated increment $\Delta\tilde{\mathbf{E}}_{n+1}^{pl}$ at time-step $n$ as:

$$\Delta\tilde{\mathbf{E}}_{n+1}^{pl} = \dot{\mathbf{E}}_n^{pl}\Delta t_{n+1} - \frac{1}{2}\ddot{\mathbf{E}}_n^{pl}(\Delta t_{n+1})^2 + \frac{1}{6}\dddot{\mathbf{E}}_n^{pl}(\Delta t_{n+1})^3 + o((\Delta t_{n+1})^4) \quad (28)$$

To compare the explicit solution with the implicit one, we expand the implicitly integrated increment $\Delta\mathbf{E}_{n+1}^{pl}$ at time-step $n$ via Taylor series:

$$\begin{aligned}
\Delta\mathbf{E}_{n+1}^{pl} &= \mathbf{E}_{n+1}^{pl} - \mathbf{E}_n^{pl} \\
&= \dot{\mathbf{E}}_n^{pl}\Delta t_{n+1} + \frac{1}{2}\ddot{\mathbf{E}}_n^{pl}(\Delta t_{n+1})^2 + \frac{1}{6}\dddot{\mathbf{E}}_n^{pl}(\Delta t_{n+1})^3 + o((\Delta t_{n+1})^4)
\end{aligned} \quad (29)$$

By subtracting Equation (29) from (28), we identify the error between the implicit solution and the assembly-free impl-exp as:

$$\begin{aligned}
\tilde{\mathbf{E}}_{n+1}^{pl} - \mathbf{E}_{n+1}^{pl} &= (\mathbf{E}_n^{pl} + \Delta\tilde{\mathbf{E}}_{n+1}^{pl}) - (\mathbf{E}_n^{pl} + \Delta\mathbf{E}_{n+1}^{pl}) \\
&= -\ddot{\mathbf{E}}_n^{pl}(\Delta t_{n+1})^2 + o((\Delta t_{n+1})^4)
\end{aligned} \quad (30)$$

We create an error bound so that the extrapolated error ($e$) is bounded for all IPs $\mathbf{X} \in \Omega$:

$$e(\mathbf{X}) = \left|\tilde{\mathbf{E}}_{n+1}^{pl} - \mathbf{E}_{n+1}^{pl}\right|_{max}(\mathbf{X}) \leq \xi \mathrm{E}^{ref} \quad (31)$$

where $\xi$ is the user-defined tolerance with respect to a model-dependent reference value $\mathrm{E}^{ref}$ and $|x_1, x_2, \ldots|_{max} = \max(|x_1|, |x_2|, \ldots)$. We now substitute Equation (30) into (31) and drop the high-order term to arrive at the error upper bound:

$$e = \left|-\ddot{\mathbf{E}}_n^{pl}(\Delta t_{n+1})^2 + o((\Delta t_{n+1})^4)\right|_{max} \simeq (\Delta t_{n+1})^2 \left|\ddot{\mathbf{E}}_n^{pl}\right|_{max} \leq \xi \mathrm{E}^{ref} \quad (32)$$

To obtain the second-order time derivative of the plastic strain, we expand its first-order derivative at time steps $n$ and $n-1$ as:

$$\begin{cases}
\dot{\mathbf{E}}_n^{pl} = \dfrac{\mathbf{E}_n^{pl} - \mathbf{E}_{n-1}^{pl}}{\Delta t_n} = \dfrac{\Delta \mathbf{E}_n^{pl}}{\Delta t_n} \\[6pt]
\dot{\mathbf{E}}_{n-1}^{pl} = \dfrac{\mathbf{E}_{n-1}^{pl} - \mathbf{E}_{n-2}^{pl}}{\Delta t_{n-1}} = \dfrac{\Delta \mathbf{E}_{n-1}^{pl}}{\Delta t_{n-1}} \\[6pt]
\ddot{\mathbf{E}}_n^{pl} = \dfrac{\dot{\mathbf{E}}_n^{pl} - \dot{\mathbf{E}}_{n-1}^{pl}}{\Delta t_n} = \dfrac{1}{\Delta t_n}\left(\dfrac{\Delta \mathbf{E}_n^{pl}}{\Delta t_n} - \dfrac{\Delta \mathbf{E}_{n-1}^{pl}}{\Delta t_{n-1}}\right)
\end{cases} \quad (33)$$

We now acquire the extrapolation error by substituting Equation (33) into (32):

$$e \simeq \frac{(\Delta t_{n+1})^2}{\Delta t_n} \left| \left( \frac{\Delta \mathbf{E}_n^{pl}}{\Delta t_n} - \frac{\Delta \mathbf{E}_{n-1}^{pl}}{\Delta t_{n-1}} \right) \right|_{max} \leq \xi \mathrm{E}^{ref} \qquad (34)$$

We compute the maximum time step at any integration point **X** while keeping the extrapolation error in bound by equating the error to the bound as:

$$\Delta t_{n+1}(\mathbf{X}) \leq \sqrt{\frac{\xi \mathrm{E}^{ref} \Delta t_n}{\left| \left( \frac{\Delta \mathbf{E}_n^{pl}}{\Delta t_n} - \frac{\Delta \mathbf{E}_{n-1}^{pl}}{\Delta t_{n-1}} \right) \right|_{max}(\mathbf{X})}} \qquad (35)$$

To bound the error in the entire domain, we set the critical step size as the minimum value of critical step sizes across all the points at the current time step $(n+1)$, that is:

$$\Delta t_{n+1}^{cri.} = \underset{\mathbf{X} \in \Omega}{\mathrm{MIN}} \sqrt{\frac{\xi \mathrm{E}^{ref} \Delta t_n}{\left| \left( \frac{\Delta \mathbf{E}_n^{pl}}{\Delta t_n} - \frac{\Delta \mathbf{E}_{n-1}^{pl}}{\Delta t_{n-1}} \right) \right|_{max}(\mathbf{X})}} \qquad (36)$$

Equation (36) provides the critical time step in our framework. We point out that $\Delta t_{n+1}^{cri.}$ depends not only on the previous time increments ($\Delta t_n$ and $\Delta t_{n-1}$) but also on the (implicit) plastic strain increments ($\Delta \mathbf{E}_n^{pl}$ and $\Delta \mathbf{E}_{n-1}^{pl}$) from the last two steps. Specifically, if the material property advances smoothly in its previous two steps, the values of plastic strain increments are close and, therefore, $\Delta t_{n+1}^{cri.}$ takes on a large value. In contrast, if the properties change abruptly (e.g., from elasticity to plasticity or vice versa), the difference between the previous two plastic strain increments is large and $\Delta t_{n+1}^{cri.}$ is reduced accordingly. Equation (36) can be further simplified as:

$$\gamma_{n+1} = \frac{\Delta t_{n+1}^{cri.}}{\Delta t_0} \approx \Delta t_n \underset{\mathbf{X} \in \Omega}{\mathrm{MIN}} \sqrt{\frac{\xi \mathrm{E}^{ref}}{\left| (\Delta \mathbf{E}_n^{pl} - \Delta \mathbf{E}_{n-1}^{pl}) \right|_{max}(\mathbf{X})}} \qquad (37)$$

where $\gamma_{n+1}$ is the ratio of time increments between the critical current time step and the (user-defined) initial step $\Delta t_0$. This simplification assumes the time increments for the last two steps are approximately the same ($\Delta t_n \approx \Delta t_{n-1}$), and their slight difference does not affect the step size at the current time point ($t_{n+1}$).

*3.2. Adaptive spatial reduction*

In addition to the adaptive temporal reduction discussed in Section 3.1, we develop an adaptive clustering-based spatial discretization method for our ROM. For any clustering-based ROM, domain decomposition generally converts the problem domain from sufficiently fine discretization (e.g., regularly shaped voxel grids from a computed tomography reconstruction or free meshes from a discretization module) into a set of interactive clusters with different shapes and sizes. Since the number of clusters is typically much smaller than the elements in the original discretization, the number of unknown variables is significantly reduced.

Clustering is an unsupervised learning technique that groups similar data points. Example clustering methods include k-means learning [37], affinity propagation [38], agglomerative clustering [39], and spectral clustering [40]. In this work, we use the k-means clustering technique which, similar to other methods, classifies mesh elements into different groups based on their feature values. The set of these features greatly affects the distinct clustering results. For example, a position-based clustering for grouping a total number of $ne$ elements in a 3D mesh uses the X, Y, and Z coordinates of each element's geometric center as features, see Table 3.

**Table 3**: **Position-based clustering data**: Elements are grouped in the k-means clustering by the 3D coordinates of their geometric centers.

|  | Features $\boldsymbol{\varphi} = [\varphi_1, \varphi_2, ..., \varphi_{nf}]^T$ | | |
|---|---|---|---|
| Data points $\boldsymbol{\zeta} = [\zeta_1, \zeta_2, ..., \zeta_{ne}]^T$ | $\varphi_1 = X$ | $\varphi_2 = Y$ | $\varphi_3 = Z$ |
| $\zeta_1 = FE_1$ | $X_1$ | $Y_1$ | $Z_1$ |
| $\zeta_2 = FE_2$ | $X_2$ | $Y_2$ | $Z_2$ |
| ... | ... | ... | ... |
| $\zeta_{ne} = FE_{ne}$ | $X_{ne}$ | $Y_{ne}$ | $Z_{ne}$ |

In k-means clustering, cluster seeds[1] are first randomly scattered in the feature space. Then, each element is iteratively assigned to the cluster whose center (i.e., mean of its members) is closest to that element. During the assignment, the cluster means are automatically updated to minimize within-cluster variances by solving the following optimization problem:

$$\boldsymbol{C} = \underset{\boldsymbol{C}}{\operatorname{argmin}} \sum_{I=1}^{k} \sum_{n \in C^I} \left\| \boldsymbol{\varphi}_n - \overline{\boldsymbol{\varphi}}_I \right\|^2 \tag{38}$$

where $\boldsymbol{C}$ refers to the created cluster centers $\boldsymbol{C} = \{C^1, C^2, ..., C^k\}$. The variables $\boldsymbol{\varphi}_n$ and $\overline{\boldsymbol{\varphi}}_I$ represent the $n^{th}$ element's feature and the feature mean of the $I^{th}$ cluster, respectively.

Naïve position-based clustering is a static spatial decomposition that is created once in an offline stage. It does not alter during runtime and results in a set of uniformly distributed material clusters with similar sizes (see [17] and Appendix A for more details). Position-based clustering may lose its data compression efficiency in the presence of nonlinear strain-softening with significant localizations. Effective clustering in such cases requires dense discretization over regions with steep gradients which can arise anywhere in the domain during the deformation (recall that the strains are constant in a cluster so capturing high gradients relies on having small clusters).

To accurately capture the softening-induced localizations in clustering, we propose a novel adaptive clustering strategy that consists of a static stage and a dynamic stage which are described in Sections 3.2.1 and 3.2.2, respectively. The implementation details of these two stages are provided in Algorithm 1.

*3.2.1. Offline static stress-informed clustering*

To assign more clusters to concentration regions in an RVE where large solution gradients exist, we propose a hierarchical clustering scheme that utilizes elemental elastic responses to guide the position-based clustering. This hierarchical scheme is used only once in the offline stage of our ROM and has two levels which firstly groups elements with the similar elastic response and then decomposes each group into some clusters[2]. The rationale behind using materials' elastic behaviors in clustering is that they are highly correlated with plastic responses. For instance, a highly stressed material point during elastic reaction is likely to obtain large plastic strains and accumulate strain concentrations during plastic deformation. Hence, our contribution in this section is the development of a hierarchical clustering scheme that creates more clusters in critical regions identified by elastic stress responses.

---
[1] The user chooses the number of seeds.
[2] While in both levels of our approach we are agglomerating elements, for clarity we reserve the terms group and cluster for the first and second levels, respectively.

At the first level, to obtain the scalar elastic stress response at each element we begin by applying a set of six orthogonal loads to the RVE. In each of the six cases, we deform the RVE by a homogeneous elastic macroscopic strain tensor and then calculate the microscale Von-Mises stress at each element's center. Next, we condense the vector containing the six elemental Von-Mises stresses (one from each load case) into a scalar via:

$$S^n = \left\|[S_1^n, S_2^n, S_3^n, S_4^n, S_5^n, S_6^n]\right\|_2 \tag{39}$$

where $S_k^n, (k = 1, \ldots, 6)$ is the Von-Mises stress at the $n^{th}$ element when the RVE is subject to the $k^{th}$ orthogonal load and $S^n$ is the L2-norm of these stresses representing the local stress intensity at an FE in the deformed RVE. The format of the resulting dataset for the stress intensity is demonstrated in Table 4 for a 3D RVE whose mesh has $ne$ elements.

Table 4: **Stress-based clustering data**: FEs are agglomerated via k-means clustering by their stress intensity values.

| Data points $\zeta = [\zeta_1, \zeta_2, \ldots, \zeta_{ne}]^T$ | Feature $\varphi = S^n$ |
|---|---|
| $\zeta_1 = FE_1$ | $S^1$ |
| $\zeta_2 = FE_2$ | $S^2$ |
| … | … |
| $\zeta_{ne} = FE_{ne}$ | $S^{ne}$ |

Once the dataset of stress intensities is built, we start to divide the elements of a meshed RVE into multiple groups (the number of groups is chosen a priori) where we assume that the elements in a group have similar stress intensity and that each group has approximately the same number of elements. We note that the elements in the same group can be anywhere in the RVE, i.e., a group can be topologically disconnected, see Figure 2.

After the groups are constructed, we then decompose each of them into multiple clusters where we assign more clusters to groups with higher stress intensities. That is, the second level of our hierarchical clustering approach aims to selectively decompose the groups such that concentration regions (marked with higher stress intensities) are discretized with more clusters. For illustration, consider Figure 2 where the elements in a generic RVE are segmented into five groups where the groups with higher stress intensities are decomposed with more clusters, and as a result, high stress-induced local phenomena in these groups can be captured more accurately.

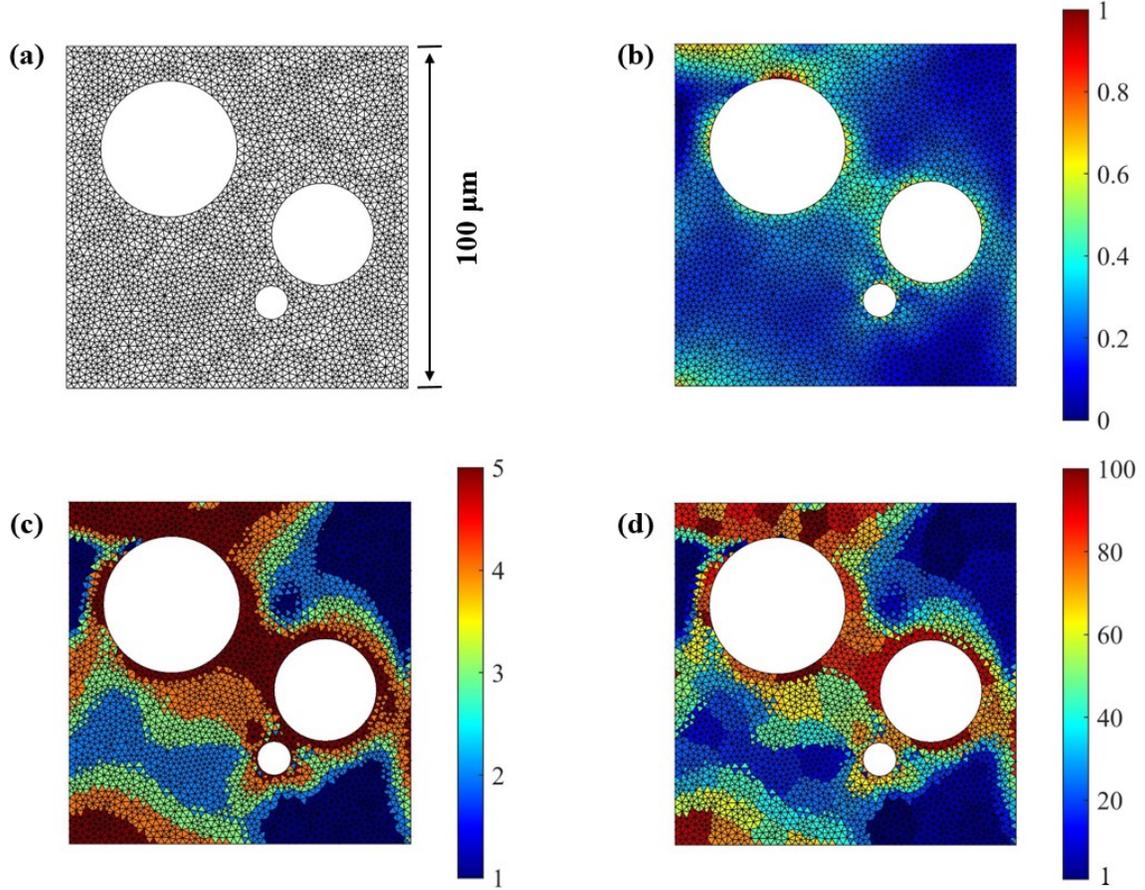

**Figure 2** Schematic 2D illustration of the offline stress-informed hierarchical clustering: **(a)** A generic 2D RVE is discretized with 5000 triangle elements. This RVE contains three circular pores with radii of 20 μm, 15 μm, and 5 μm. **(b)** Element stress intensities are computed via offline orthogonal loads where intensity values are scaled to [0, 1] to aim in visualization. **(c)** RVE elements are firstly agglomerated into five groups based on elemental intensity values where each group has 1000 elements. The intensity monotonically increases from group 1 to group 5. **(d)** Elements are then classified into a total of 100 clusters where the groups 1, 2, 3, 4, and 5 contain 10, 15, 20, 25, and 30 clusters, respectively.

We introduce the clustering split factor ($s_f$), to better control the number of clusters in each group. Specifically, we first sort the groups in an ascending order based on their stress intensity values. Then, we compute the number of clusters in a group via a power function whose exponent is $s_f$:

$$N_{g_i}^{cl} = \frac{\theta_{g_i}}{\sum_{i=1}^{N_g} \theta_{g_i}} N^{cl} \tag{40}$$

$$\theta_{g_i} = \left(\frac{I_{g_i}}{N_g}\right)^{s_f} \tag{41}$$

where $N_{g_i}^{cl}$ is the number of clusters in the group $g_i$, $\theta_{g_i}$ is the fraction of total clusters assigned to $g_i$, $I_{g_i}$ is the group's sorted index in the ascending order of stress intensities, $N^{cl}$ is the prescribed total number of clusters, and $N_g$ is the total number of groups. If the computed $N_{g_i}^{cl}$ is decimal, we round it up to the nearest integer and adjust the number of clusters in other groups accordingly such that $N^{cl} = \sum_{i=1}^{N_g} N_{g_i}^{cl}$.

We note that $s_f$ is defined by the user and represents the contrast of cluster densities across the groups. This contrast is illustrated in Figure 3 where the cluster numbers per group are controlled by adjusting $s_f$. We observe that a large $s_f$ promotes increasing cluster numbers in groups with high-stress intensities. In contrast, a small $s_f$ lowers the sensitivity of cluster density to stress intensity. Specifically, when $s_f = 0$ the numbers of clusters in each group are the same, and when $s_f > 0$ more clusters are generated in groups with higher average intensity values. Also observing that an excessively large split factor can dramatically reduce clusters in the groups with low-stress intensities. For example, when $s_f = 10$ most clusters are created in the group with the highest intensity, and many groups with low intensities only contain one cluster. One cluster cannot capture solution gradients, and if local phenomena occur in those regions during the online stage (overlooked in offline elastic tests), the lack of sufficient clusters will result in dramatic local inaccuracy. Therefore, we do not recommend large $s_f$ and we set the default value to 1.0 which makes the number of clusters in each group to be linearly proportional to the sorting index of stress intensities. We illustrate the effects of $s_f$ on the accuracy of our ROM in Section 4.2.

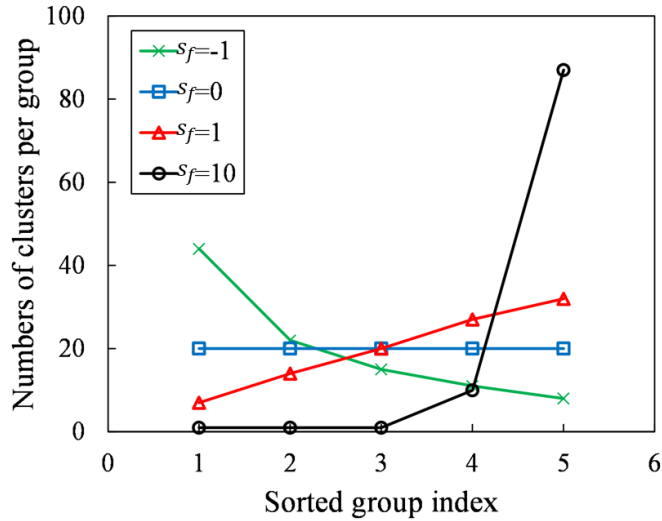

**Figure 3 Effects of split factor:** The numbers of clusters in groups with high-stress intensities (marked by high sorted indices) rapidly increase as a positive $s_f$ grows. In contrast, when n is negative, more clusters are assigned to non-critical groups with smaller intensity values. This figure is generated by distributing 100 clusters between five groups.

We point out that the total number of clusters, $N^{cl}$, is determined by the user and then they are distributed across the groups by Equation (40). If $N^{cl}$ is too small, too few clusters in a group may not sufficiently represent sharp solution gradients in a group. Hence, we require the computed cluster number from Equation (40) to be at least as large as $N_{g_i}^{Cl*}$ which is the smallest number that well-behaved clustering requires, that is:

$$N_{g_i}^{cl} = \max(N_{g_i}^{cl}, N_{g_i}^{Cl*}) \tag{42}$$

There are multiple techniques to compute $N_{g_i}^{Cl*}$ including gap statistics [41], elbow method [42], Silhouette coefficient [43], and Calinski-Harabasz index [44]. We adopt the Calinski-Harabasz index in this work which is computed as:

$$CH_k = \frac{V_b}{V_w} \times \frac{N_{elem}^{g_i} - k}{k-1} \tag{43}$$

where $CH_k$ is the Calinski-Harabasz index of $k$ clusters, $V_b$ is the overall between-cluster variance, $V_w$ is the overall within-cluster variance, and $N_{elem}^{g_i}$ is the number of elements in the $i^{th}$ group. The overall between-cluster variance and the overall within-cluster variance are defined as:

$$V_b = \sum_{j=1}^{k} N_{elem}^{cl_j} \|m_j - m\|^2 \tag{44}$$

$$V_w = \sum_{j=1}^{k} \sum_{m_e \in cl_j} \|m_e - m_j\|^2 \tag{45}$$

where $N_{elem}^{cl_j}$ is the number of elements in the $j^{th}$ cluster of the $i^{th}$ group and $m_j$ denotes the centroid of the $j^{th}$ cluster. $m$ and $m_e$ are the overall mean of feature values for all elements in the $i^{th}$ group and feature of one element, respectively. Since a good clustering has a large between-cluster variance and a small within-cluster variance, i.e., a large Calinski-Harabasz index value, the optimal cluster number for the $i^{th}$ group can be determined by the following optimization problem:

$$N_{g_i}^{Cl*} = \underset{k}{\mathrm{argmax}}(CH_k) \tag{46}$$

where the applied cluster number for the $i^{th}$ group in k-means clustering is obtained by comparing $N_{g_i}^{Cl*}$ with $N_{g_i}^{cl}$, see Equations (40) and (42).

The hierarchical clustering is performed once in the offline stage of our ROM. In the next section, we augment this static offline clustering with a dynamic counterpart that evolves with iterative solutions of the nonlinear equilibrium equations of plasticity and damage.

### 3.2.2. Online dynamic clustering

We propose an online dynamic spatial decomposition method to accommodate the evolution of equilibrium equations during the initiation and propagation of strain hardening and softening. The dynamic method augments the static offline clustering of Section 3.2.1 with online refinements over critical regions in an RVE as it is deformed. In a multiscale simulation, the dynamic clustering is performed for all the RVEs associated with the macro-IPs.

While an RVE is deformed, we check a dynamic clustering initiation condition at each load increment. If the condition is not met, we solve the micro-equilibrium equations based on the existing clustering. Otherwise, we activate the dynamic refinement and calculate an error indicator that selects the "parent" clusters that must be decomposed into multiple "children" clusters. This decomposition relies on a remeshing procedure for dividing the clusters and transferring the state variables between the parent and children clusters. We stop the online decomposition once the total number of child clusters exceeds the user-defined total number of clusters ($N_{user}^{cl}$), that is:

$$\sum_{t=1}^{n+1} N_t^{cl} \leq N_{user}^{cl} \tag{47}$$

where $N_t^{cl}$ is the number of new clusters created at time step $t$. Details of each step are discussed below.

In this work, we assume adaption starts and continues as long as the total number of existing clusters is less than the prescribed upper limit that is chosen by the user. If this condition is satisfied, we select the parent clusters that must be refined based on their equivalent plastic strain values. Specifically, we decompose the clusters whose plastic strains are either the highest in the RVE (spatial anomaly) or quite different from the plastic strains of the neighboring clusters (spatial discontinuity). We note that our adaption scheme is integrated with DCA [17] whose clusters are topologically connected (unlike the approach developed in [28]). Consequently, finding a cluster's neighbors and computing its spatial discontinuity is simply done by checking the connectivity of agglomerated elements followed by comparing their equivalent plastic strains.

The clusters with spatial anomalies are likely located in a strain softening region while those with spatial discontinuities tend to be at the boundary of damage bands surrounded by fractured and intact elements. We consider both cases as indicators of concentration regions which require higher degrees of interpolations (i.e., more cluster decomposition) for capturing solution gradients accurately.

Compared to existing adaptive methods such as the one in [45], the novelty of our approach is that we augment the two spatial softening indicators discussed above with a temporal metric to choose the proper time steps for spatial refinements. Specifically, we systematically integrate our dynamic clustering with the new AAF-IE scheme proposed in Section 0 to identify the transition time of material properties between elasticity and plasticity. The identification of the critical time instances helps to improve fracture modeling since increasing spatial interpolations at material transition stages are crucial to predict accurate strain states that precede plasticity and damage.

Once the parent clusters are selected at critical transition time instances, we refine them by re-meshing which controls the number of children clusters created at each step. This re-meshing, similar to the h-adaptivity of FEM, decomposes a parent cluster into multiple small clusters while maintaining the original boundaries. Since the parent clusters are in concentration regions with high solution gradients, this decomposition improves the solution accuracy by enriching spatial interpolations.

We divide each parent cluster by the position-based clustering which agglomerates material points based on their coordinates to multiple child clusters with topologically connected boundaries. For simplicity, we assume each parent cluster is decomposed into two child clusters, and neither of the new clusters can be further divided. Furthermore, we use the AAF-IE metric in Equation (37) to approximate the number of clusters that must be added at the current time instance $t_{n+1}$ as a function of the time increment $\Delta t_n$ and the plastic strain increments ($\Delta \mathbf{E}_n^{pl}$ and $\Delta \mathbf{E}_{n-1}^{pl}$) from previous steps, that is:

$$N_{n+1}^{cl} = N_0^{cl} / \gamma_{n+1} \approx N_0^{cl} / \MIN_{\mathbf{X} \in \Omega} \Delta t_n \sqrt{\frac{\xi \mathrm{E}^{\mathrm{ref}}}{\left|(\Delta \mathbf{E}_n^{pl} - \Delta \mathbf{E}_{n-1}^{pl})\right|_{\max}(\mathbf{X})}} \qquad (48)$$

where $N_0^{cl}$ is the prescribed number of new clusters when the adaption step starts.

The underlying idea of Equation (48) is to have the number of new clusters be inversely proportional to the AAF-IE's error that estimates the variation of material responses amid transitions. As a result, we provide the transition phase (when extrapolation errors are expected, see Section 3.1) with smaller time steps and finer spatial discretization. In Section 4.2 we demonstrate the advantage of dynamic clustering over a simple approach where a fixed number of clusters are added per load increment.

A major advantage of our dynamic clustering strategy is that parent and their child clusters are hierarchically related. That is, upon refinement, child clusters inherit the same history-dependent state variables from their parents and we only need to locally adjust the reduced stiffness matrix in DCA via the incremental assembly by Equation (24). The implementation details of our dynamic and static clustering methods are outlined in Algorithm 1.

---

**Algorithm 1** Structure of our adaptive clustering approach
---
1: **procedure** Develop microscopic fully adaptive clusters for highly localized phenomena
2:     ▷Offline stage: stress-informed static clustering
3:     **if** $t \leftarrow 0$     ▷At the initial time step
4:         Read the microstructure's FE mesh and material properties
5:         Apply six orthogonal loadings via elastic macro-strains and compute stress norm for each element
6:         Agglomerate micro-elements to groups based on element stress norm value
7:         Calculate the number of cluster seeds in each group according to their average stress values
8:         Check if the number of cluster seeds per group is optimal by the Calinski-Harabasz criterion
9:         **if** Calinski-Harabasz criterion is satisfied, **then**
10:             Generate the static stress-informed clusters for each group by the optimal seed numbers
11:         **else**
12:             Adjust group seed numbers and re-check Calinski-Harabasz criterion
13:     **end if**
14:     ▷Online stage: dynamic clustering
15:     **if** $t > 0$     ▷Increment the time steps in online computing
16:         **for** $j \leftarrow 0, N$ **do**     ▷Loop over $N$ load steps
17:             Read macroscopic deformation information
18:             Solve micro-equilibrium solutions
19:             Check clustering adaption criterion
20:             **if** clustering adaption is satisfied, **then**
21:                 Identify target decomposing clusters via spatial error indicator
22:                 Determine critical adaption time steps via temporal error indicator
23:             **else**
24:                 Proceed to the next load step
25:             **end if**
26:             Estimate each group's new cluster numbers via temporal error estimator
28:             Compare the accumulated new cluster number to the prescribed upper limit
29:             **if** the current cluster number < upper limit, **then**
30:                 Decompose target clusters and generate new clusters
31:                 Transfer state variables from target clusters to their child clusters
32:                 Update reduced stiffness matrix on entries associated with cluster refinement
33:             **else**
34:                 Proceed to the next load step
35:             **end if**
36:         **end for**
37:     **end if**
38: **end procedure**

---

### 3.3. Workflow of our multiscale damage model

As demonstrated in Figure 4 our model starts with the offline stress-informed clustering (see Section 3.2.1) which solves the multiscale nonlinear analysis by incrementing load steps to estimate the macroscale deformation gradients and microscale effective responses. This estimation is done via a hybrid integration scheme which avoids softening-induced solution divergence (see Sections 2.1 and 2.4). In this process, we perform online adaptive temporal adjustment (see Section 3.1.2) and spatial decomposition (see Section 3.2.2) to improve the analysis accuracy in regions where strain-localized softening behaviors appear.

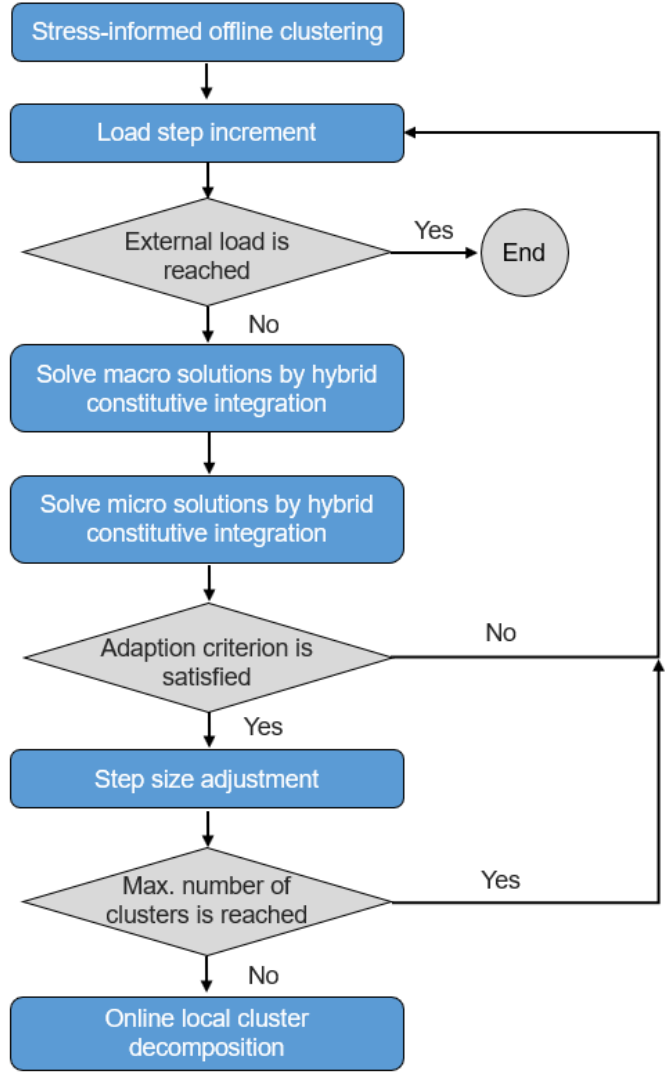

**Figure 4 Flowchart of our model:** Our approach has offline and online stages where the latter stage allocates clusters to regions where high strain concentrations appear during the simulation.

## 4. Numerical experiments

In this section, we study the effect of porosity on the softening behavior of metallic components made of aluminum alloy A356. We consider the alloy as the primary material phase and the pores as the secondary phase. Except for porosity, other polycrystalline microscopic features such as grain boundaries are out of the scope of this study. We perform the spatial and temporal adaptive reductions introduced in Sections 3.1 and 3.2 on the primary material phase in all micro-, macro-, and multi-scale simulations. The values of elastic modulus ($M$) and Poisson's ratio ($v$) of A356 are given as:

$$M = 5.70\text{e}4 \text{ MPa}, \quad v = 0.33 \tag{49}$$

A356's elastoplastic hardening behavior is assumed to be isotropic and follow an associated plastic flow rule with the Von-Mises yield surface defined as:

$$S \leq S_Y \left( \bar{E}^{pl} \right) \tag{50}$$

where $S$ is the Von-Mises equivalent stress and the yield stress $S_Y$ is governed by a predefined hardening law that depends on the equivalent plastic strain $\bar{E}^{pl}$. The material's hardening behavior is assumed as piecewise linear as demonstrated in Figure 5. To model strain softening, the alloy is modeled as a ductile metal with the fracture strain ($E_f$) and fracture energy ($G_f$) given as:

$$E_f = 6.67\text{e-}2, \quad G_f = 1.92\text{E4 N/m} \tag{51}$$

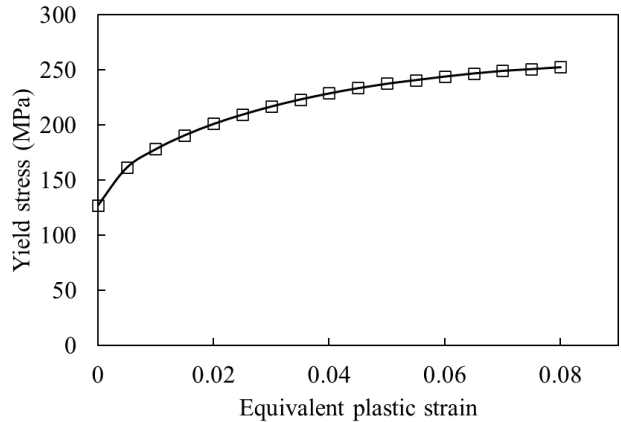

**Figure 5 Hardening behavior of A356:** We use a piecewise linear hardening behavior.

As discussed in Section 3, our ROM framework employs temporal and spatial reductions to accelerate damage simulations. Correspondingly, in Section 4.1, we study a macroscopic 3D plate to demonstrate the benefits of the proposed AAF-IE approach discussed in Section 3.1 which achieves temporal reduction by reducing required time steps in intricate nonlinear analyses. Next, in Section 4.2, we test the adaptive spatial clustering techniques discussed in Section 3.2 to deform 3D microstructures under complex loading conditions. Finally, in Section 4.3 the two reduction techniques are integrated within the first-order computational homogenization scheme to quantify the effects of spatially varying micro-porosity on macro-component damage behaviors. In all experiments, the accuracy of the proposed ROM is verified against direct numerical simulations (DNS) based on FEM.

The proposed method is implemented in Matlab. All macro- and micro-scale simulations in Sections 4.1 and 4.2 are conducted on a 64-bit Windows desktop with the following hardware: four Intel i5-8250U CPU cores running at 1.8 GHz with 16 GB installed RAM. The concurrent multiscale simulation in Section 4.3 is performed on a high-performance cluster with 30 CPU cores (AMD EPYC processor running at 4.1 GHz) and 192 GB RAM.

*4.1. Macroscale experiments*

We test the AAF-IE method on the macroscale 3D plate shown in Figure 6 (a). Due to symmetry, we deform only one-quarter of the model by applying a Dirichlet boundary condition ($\bar{u} = 0.8$ mm) on the top surface and symmetric boundary conditions on the left and bottom surfaces, see Figure 6(b). To demonstrate the efficiency of AAF-IE, we first consider an elastoplastic deformation with hardening but without fracture. To obtain a benchmark solution, we mesh the model with 6000 linear tetrahedron elements and compute the solution via an implicit scheme. The obtained equivalent plastic strain distribution is shown in Figure 6(c) where the elements with the highest strain values are observed near the surface of the middle hole.

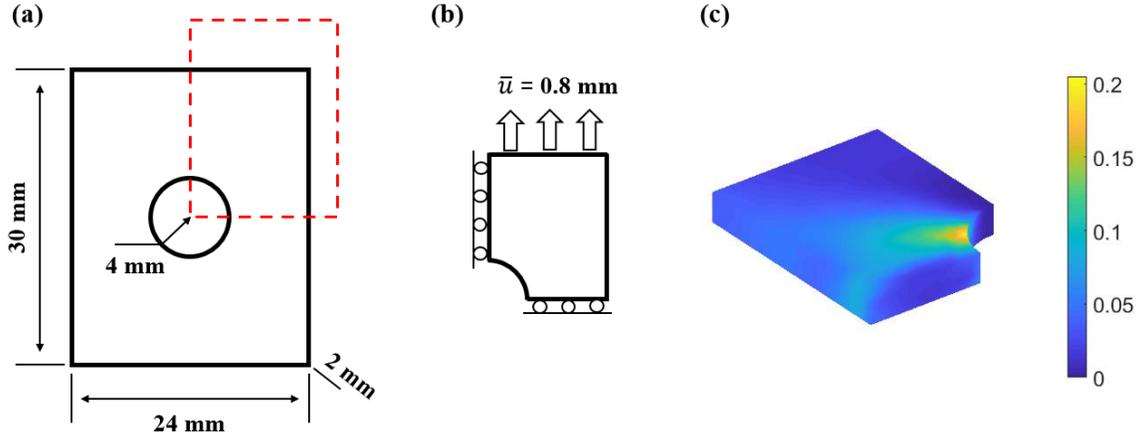

**Figure 6 Macroscale model: (a)** Geometry and the dimensions of the 3D plate which has a hole in its center. Due to symmetry, only the red dashed part is modeled. **(b)** Boundary conditions are applied to the studied model. **(c)** Distribution of the equivalent plastic strain after work hardening (no damage, obtained via DNS).

The AAF-IE's load-displacement response is demonstrated in Figure 7(a) where the implicit solution is included as a benchmark. Since the current state variable (increment of plastic strain tensor) in an impl-exp scheme is linearly extrapolated from its previous values, see Section 3.1, prediction accuracy is affected by its step sizes at material transition. In particular, the impl-exp scheme's extrapolation errors are significant when using large steps in the transition. For example, in the constant-step assembly-free impl-exp (CAF-IE) schemes, decreasing the number of steps from 200 to 50 results in larger time steps and higher extrapolation errors, see Figure 7(a). On the contrary, the extrapolation error of our AAF-IE scheme is reasonably small even though we use much fewer time steps. This is because our adaptive schemes automatically adjust timestep sizes such that smaller steps are utilized in the transition of property phases while large steps are used at other non-critical times.

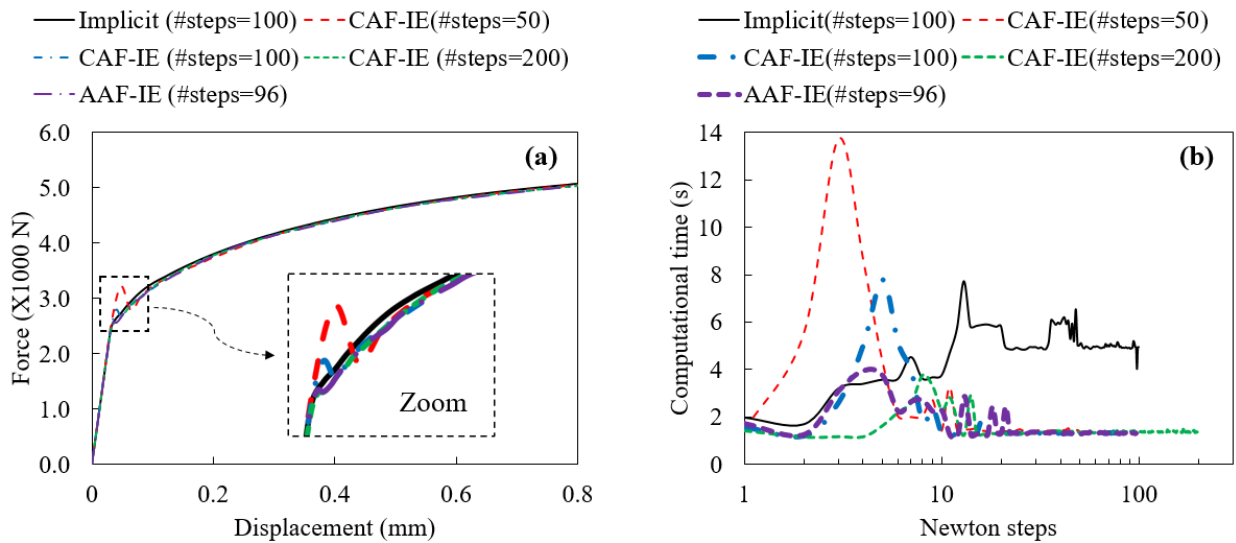

**Figure 7 Accuracy and efficiency assessment of IE schemes: (a)** Comparison of load-displacement responses for a strain-hardening simulation between the implicit solver, CAF-IE with different constant steps, and AAF-IE. **(b)** The computational time of each Newton step for different schemes. A close-up comparison of CAF-IE schemes is shown without other solutions.

We note from Figure 7(a) that during the material transition phase fewer constant steps of CAF-IE result in larger step sizes and, as a result, exhibit higher extrapolation errors (e.g., comparison of CAF-IE with 50 steps and 200 steps). With a larger extrapolation error, more iterations are required in CAF-IE towards convergence at each step which, in turn, results in higher computational time at each Newton step as shown in Figure 7(b). Comparatively, our AAF-IE adaptively reduces its step size to lower the required number of iterations and computational time per step. After the transition phase, extrapolation errors are greatly reduced such that fewer iterations and smaller computational time are required at each step. Comparatively, for the implicit scheme, its computational cost per iteration is much higher than the other schemes because it requires the underlying stiffness matrices to be re-assembled at every iteration.

Note that the number of steps in CAF-IEs (i.e., 50, 100, 200 in Figure 7(b)) are prescribed by the user where smaller numbers result in longer per-step computational costs since the steps are larger and involve more iterations. Meanwhile, the number of steps in AAF-IE (i.e., 96) is automatically determined by the algorithm to achieve the same level of solution accuracy as the CAF-IE with 200 steps, see Figure 7(a). The reason AAF-IE needs fewer steps than CAF-IE is that AAF-IE can adjust its temporal discretization where steps are enlarged during non-critical time instances and, as a result, the number of steps decreases.

We now demonstrate the performance of AAF-IE in the presence of softening by comparing it against two fracture simulations that use explicit and co-simulation schemes, see Figure 8(a). The explicit solver is conditionally stable but is quite slow since sufficiently small steps are required to ensure stability. The co-simulation approach applies the implicit and explicit solvers sequentially to simulate strain hardening and softening, respectively. It is more efficient than the pure explicit solver, but its efficiency depends on the fraction of deformation that includes softening.

The load-displacement curves obtained from different methods are illustrated in Figure 8(a) where we observe that our AAF-IE and CAF-IE with 200 constant steps achieve very similar results. However, the solutions obtained via explicit and co-simulation methods indicate significant fluctuations in both hardening and softening. Reducing such erroneous fluctuations often requires applying fictitious damping forces which can result in unphysical solutions if they are excessive and implemented improperly. In addition, it is noteworthy that our AAF-IE (i.e., 100 steps) only requires half of the time steps as that of CAF-IE (i.e., 200 steps), indicating a dramatic temporal reduction.

We compare the total computational time of the four different solvers in Figure 8(b) where we observe that CAF-IE and AAF-IE are significantly faster than the explicit and co-simulation procedures. This computational efficiency is primarily because both CAF-IE and AAF-IE are unconditionally stable and hence their enlarged steps reduce the overall simulation time. We also note that AAF-IE converges to the same solution as CAF-IE but with almost half of the steps and a 59% reduction in costs. Compared to the explicit solver widely applied in fracture mechanics, our AAF-IE scheme achieves a more accurate (smooth) solution with an acceleration factor of 11.4. We, therefore, apply AAF-IE to the following studies.

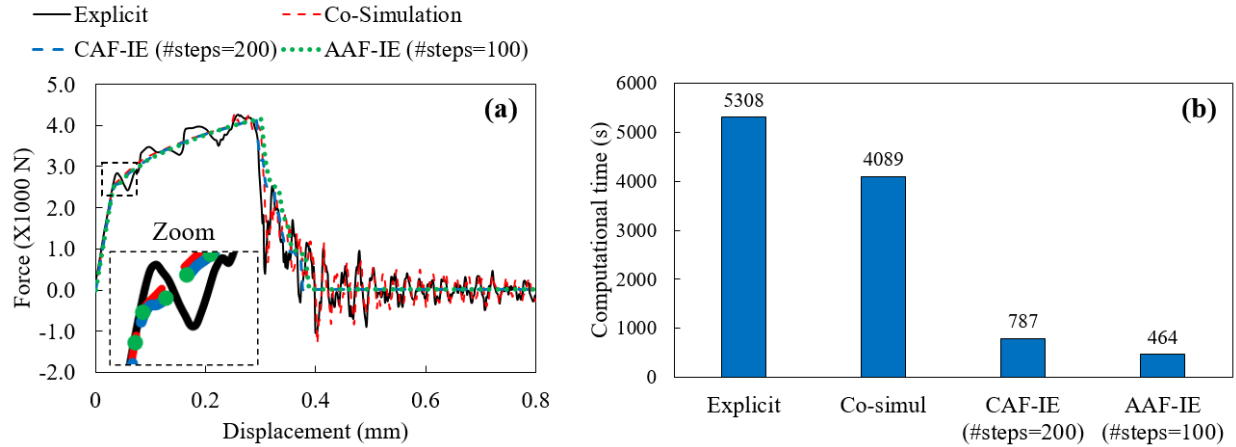

**Figure 8 Fracture simulations via different solvers: (a)** Comparison of load-displacement responses between explicit solver, co-simulation scheme, IE solver with constant steps, and our AAF-IE solver. **(b)** Comparison of total computational time between different solvers.

We prevent the softening solutions from mesh dependency by applying a non-local function (see Section 2.3) to the macroscale model in Figure 6. Specifically, we assume the strain localization band has a width[3] of 1.5 mm and then test for mesh independency by discretizing the model with four mesh levels and comparing the resulting load-displacement responses. As shown in Figure 9, the softening curves converge as the number of elements increases. With 13000 elements, mesh independency is observed for both post-failure load-displacement responses and fracture energies (i.e., the area under the load-displacement curve after damage initiation).

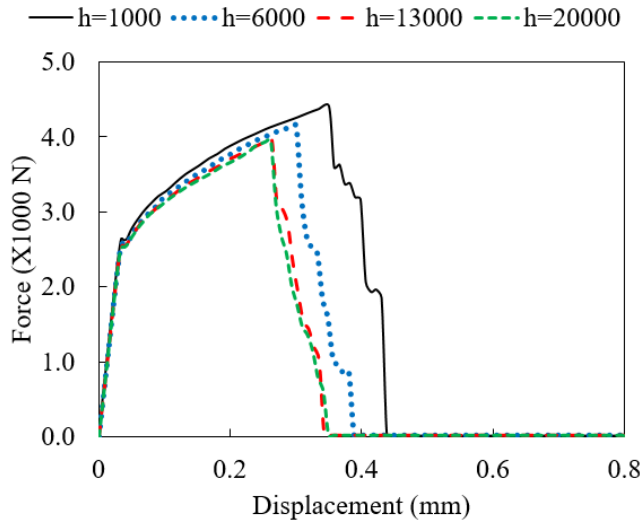

**Figure 9 Mesh independency study on history variables:** Load-displacement responses converge as the number of elements (h) increases.

We can further verify mesh independency by comparing the equivalent plastic strain distributions and damage patterns across the four mesh levels, see Figure 10. As the number of elements increases in Figure 10(a)-(d), we observe similar strain localizations. In particular, the direction of the softening band in each plot is orthogonal to the direction of the applied Dirichlet boundary condition. This band marks the damaged elements and extends from the inner hole to the outer surface. During damage evolution, the elements inside the band accumulate significant plastic strain and so their damage variables possess much higher values than their surrounding elements, see Figure 10(e)-(h). This observation explains the striking similarity between the strain localization bands (top row) with the fracture bands (bottom row) in Figure 10.

---
[3] A more realistic softening bandwidth value is typically determined via experiments or high-fidelity simulations.

It is noteworthy that in Figure 10(e), the mesh is so coarse that the element sizes are almost the same as the prescribed fracture band width. In this scenario, the non-local function in Equation (14) has little effect and the fracture band contains one single layer of damaged elements. However, as discretization become increasingly finer in Figure 10(f)-(h), the non-local function achieves convergent softening solutions by averaging the damage variables of elements within the prescribed width of fracture bands. Across these cases, we also observe that multiple layers of damaged elements are contained within the fracture bands which indicates that the macroscale damage patterns are independent of mesh sizes and produce consistent fracture energies.

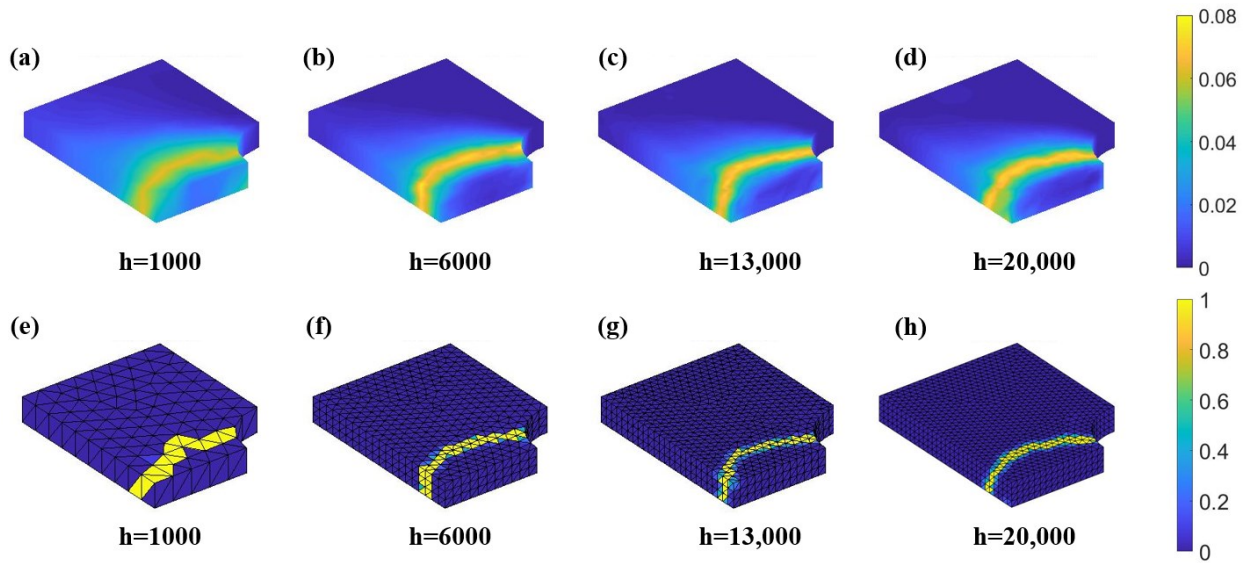

**Figure 10 Macroscale mesh independency study of field variables: (a-d)** Distributions of equivalent plastic strains which converge as the number of elements increases. **(e-f)** Damaged elements and the distributions of fracture bands are independent of the mesh.

*4.2. Microscale experiments*

In this section, we integrate the temporal and spatial adaption techniques into our ROM whose performance is then tested on various porous microstructures. In all experiments, microscale pores are modeled as prolate ellipsoids with two identical minor axes. We use the following four descriptors to characterize the morphologies and spatial distribution of pores in a microstructure: pore volume fraction ($V_f$), number of pores ($N_p$), aspect ratio between major and minor axes ($A_r$), and the average spatial distance between the two nearest pores ($\bar{r}_d$, units in μm). In addition, we simulate the microscale damage evolutions via the stabilized micro-damage model proposed in [31] (see also Appendix B).

*4.2.1. Effect of different clustering methods*

In this experiment, we apply our ROM to the microstructure in Figure 11(a) which has two identical spherical pores located on the neutral z-plane. The dimensions of the microstructure's cross-section at the neutral z-plane are shown in Figure 11(b). Figure 11(c) shows the distribution of the equivalent plastic strain obtained via DNS (FEM with 33000 elements) when this microstructure is subject to the macroscopic deformation gradient in Equation (52).

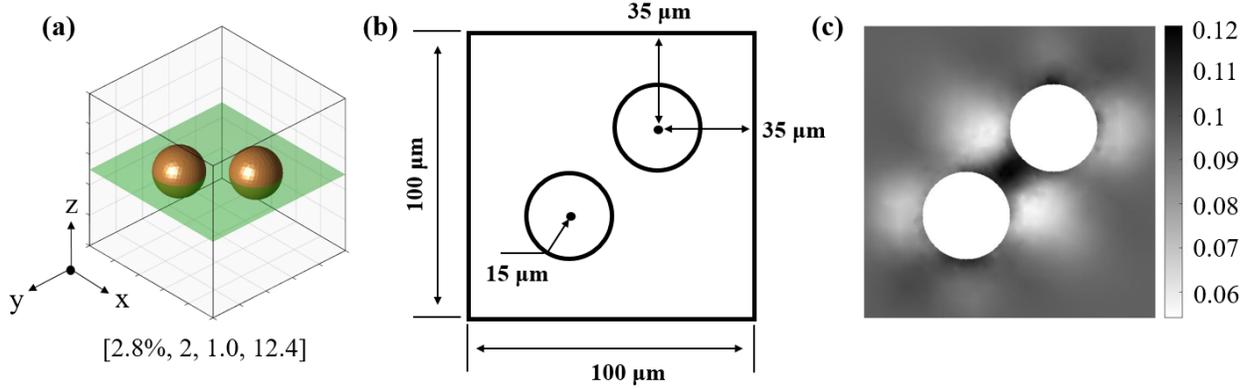

**Figure 11 A simple 3D microstructure with two pores: (a)** The microstructure contains two identical spherical pores on the neutral z-plane. **(b)** Dimension of the microstructure (on the neutral z-plane). **(c)** Distribution of the equivalent plastic strain via DNS.

$$\mathbf{F} = \begin{bmatrix} 1.1 & 0 & 0 \\ 0 & 0.95 & 0 \\ 0 & 0 & 0.95 \end{bmatrix} \tag{52}$$

We start by checking the sensitivity of the microscale damage solutions to mesh size in DNS. To this end, we discretize the microstructure by four different meshes which have 2000, 8000, 15000, and 33000 elements. By comparing their homogenized stress-strain responses in Figure 12(a), we find that the microscale mesh dependency is well controlled by using the fracture energy-constrained microscale damage model (see Section 2.3.2 and Appendix B). Specifically, as the number of elements increases, the effective responses converge to the solution with 33000 elements. To emphasize the importance of the prescribed fracture energy on damage responses, we add two curves in Figure 12 that correspond to a simulation with larger fracture energy ($G_f^* = 1.92e6$ N/m), see Equation (51) for $G_f$. As these curves indicate, the post-damage behavior and fracture energy are quite sensitive to the prescribed $G_f$. Particularly, a larger fracture energy enables material points to withstand higher external loads before fracture.

We now compare DNS with our ROM with the position-based clustering. The results are provided in Figure 12(b) and indicate that as the number of clusters increases, the predicted ultimate tensile strength (UTS) and fracture toughness approach to those of DNS. Specifically, with 1600 clusters, ROM's homogenized response is identical to that of the DNS. It is noteworthy that the number of clusters represents the levels of spatial discretizations. A small cluster number generally results in coarse discretization with diffusive (low) plastic strain fields at each cluster. As the magnitude of the local plastic strain is smaller than DNS, ROMs exhibit delayed fracture initiations, higher UTS, and larger material toughness. This delayed damage response is gradually resolved as we increase cluster numbers. We also note that the ROM solutions are obtained by the microscale damage model (see Appendix B) which uses the same $G_f$ as in DNS. Similar to above, the ROM with a larger fracture energy $G_f^*$ is able to endure higher macro-strain before softening, and thus exhibit delayed damage behavior.

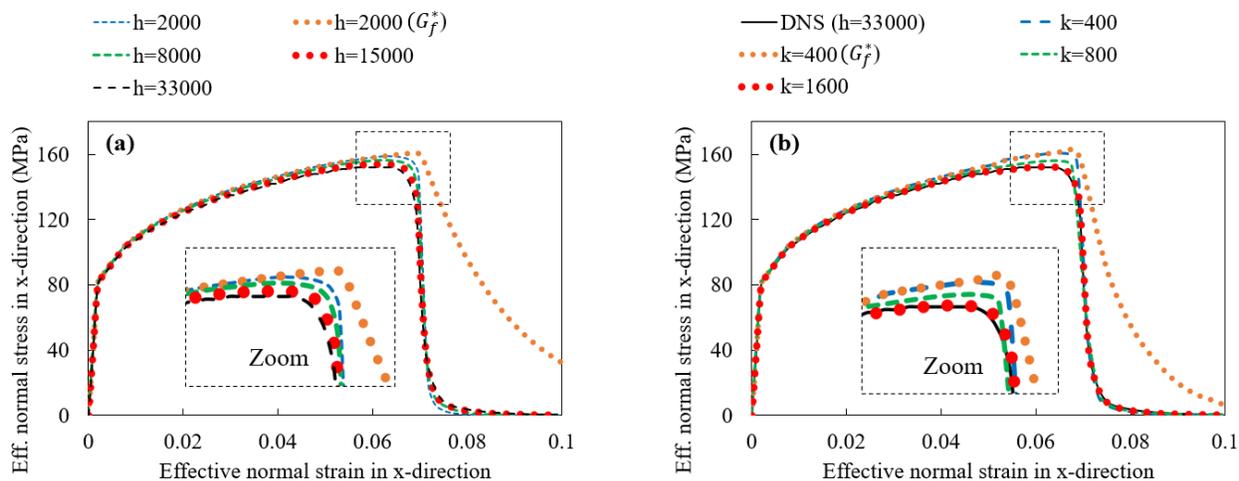

**Figure 12 Homogenized results by DNS and ROM: (a)** Mesh independency study in DNS (h: number of elements) where the evolution of the damage variable is controlled by fracture energy $G_f$. Fracture behaves very distinctively when using $G_f^*$. **(b)** Results of ROM homogenizations via a position-based clustering (k: number of clusters) where the impacts of $G_f$ is demonstrated by providing the curve corresponding to $G_f^*$ for k=400.

We next demonstrate the efficacy of the stress-informed clustering by preprocessing the same RVE with six offline orthogonal loads as discussed in Section 3.2.1. The resulting Von-Mises stresses are used to compute the elemental stress intensity scalar whose spatial distribution is demonstrated in Figure 13(a). Following our hierarchical approach, we divide the elements into three groups where each group has approximately the same number of elements, see Figure 13(b). We choose three different values for the total number of clusters (i.e., $k = 400, 800,$ and $1600$) and then follow the procedures outlined in Section 3.2.1 to assign these clusters to the three groups. After hierarchical clustering (which is in the offline stage and done only once), we predict the fracture behavior in the online stage via our ROM. The results are summarized in Figure 13(c) and indicate that, with the same number of clusters, stress-informed clustering provides higher accuracy than the naïve position-based clustering in Figure 12(b). Specifically, in Figure 13(c), we note that when the number of clusters increases to 800, the ROM's solution converges to that of DNS.

The clustering split factor in Equation (40) is an important parameter that affects the results of our hierarchical clustering. To demonstrate its impacts, we compare the solutions of ROMs with the same 400 clusters but different values of $s_f$ in Figure 14. As discussed in Section 3.2.1, a positive $s_f$ promotes more clusters in the groups with high stress intensities, while a negative $s_f$ works oppositely. For instance, $s_f = 1.0$ results in 67, 133, and 200 clusters in groups 1, 2, and 3, respectively, while $s_f = -1.0$ genrates 219, 109, and 72 clusters in those groups. Comparing the effective results from the two $s_f$, we observe a significant improvement when assigning more clusters in critical regions ($s_f = 1.0$). Therefore, we choose the default value of the heuristic split factor as 1.0 in our following tests.

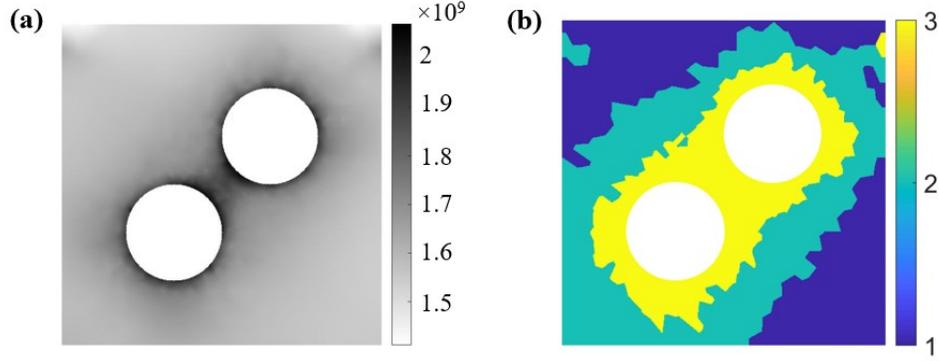

**Figure 13 Stress-informed clustering: (a)** Distribution of stress intensities which is computed from the six orthogonal offline loadings. **(b)** Elements are agglomerated into three groups according to their stress intensities where each group has approximately the same number of elements. Both figures are plotted on the neutral z-plane of the 3D RVE in Figure 11(a).

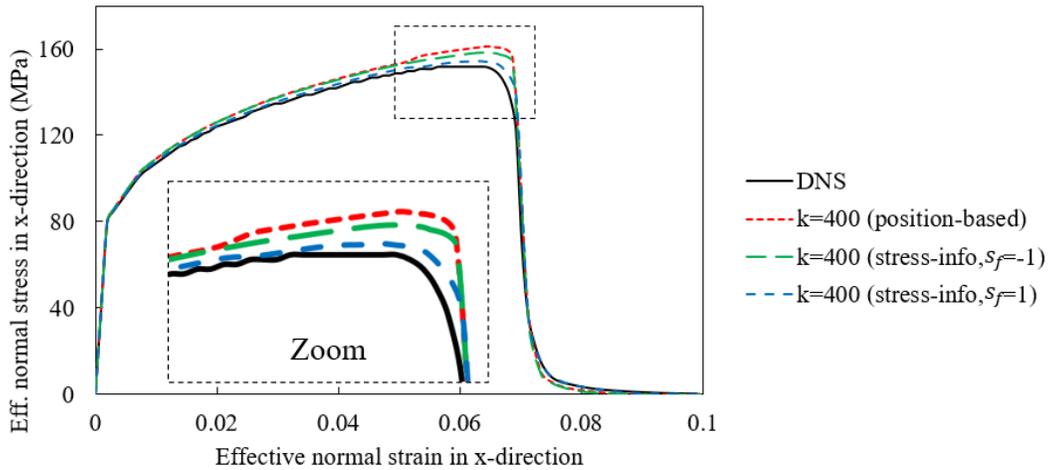

**Figure 14 Effective responses of stress-informed clustering:** Compared to position-based clustering results, stress-informed solutions are improved using the same number of clusters (k=400). A higher value of clustering split factor in the stress-informed method results in more accurate solutions.

To further distinguish the position-based and static stress-informed clusterings, we compare their cluster distributions and equivalent plastic strains in Figure 15 where the RVE is decomposed by 400 and 1600 clusters. Compared to the position-based method, the stress-informed approach allocates more clusters to the highly-stressed regions (around the bridge region connecting the two pores) and fewer clusters to the non-critical regions, compare the top row in Figure 15(a) and (c) with Figure 15(b) and (d). As a result of such distributions, the stress-informed method produces large (small) clusters in non-critical (stressed) regions which increase solution accuracy as small clusters enable capturing high gradients. This improvement is illustrated in the bottom row of Figure 15 where stress-informed clustering captures strain concentrations better than the position-based counterpart. We note that, while the results based on these two clustering methods differ, they are both quite close to DNS results (e.g., compare the equivalent plastic strains in Figure 15 with DNS results in Figure 11(c)).

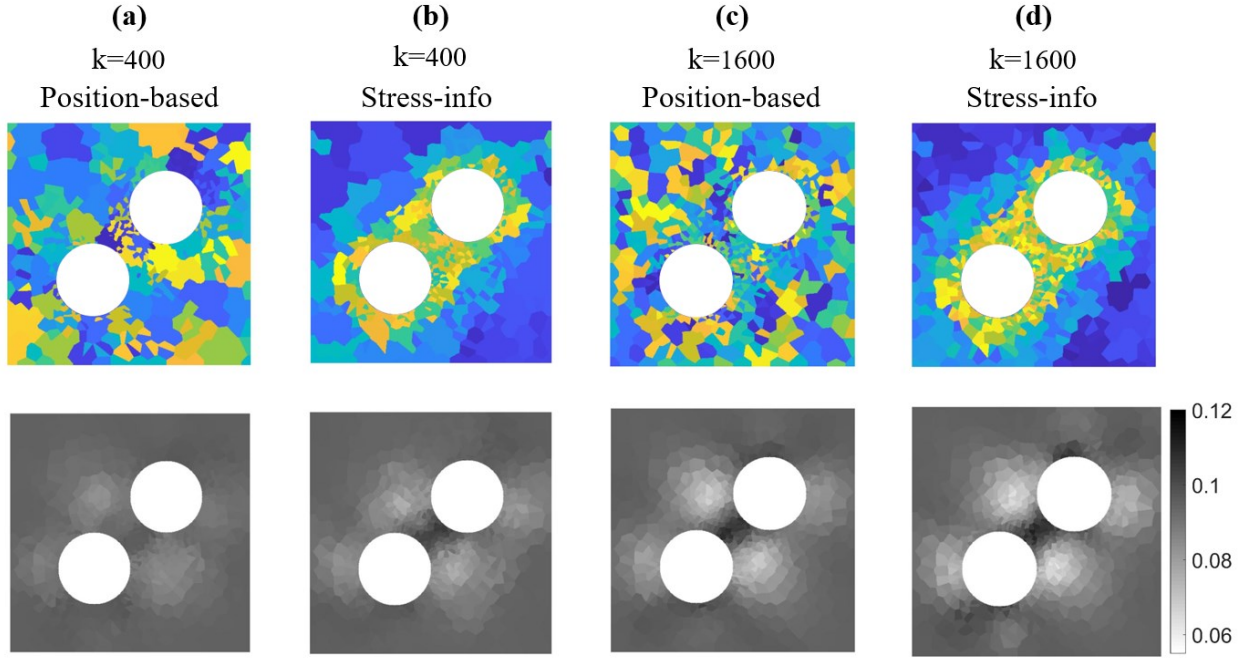

**Figure 15 Comparisons of clustering and effective plastic strains:** The top and bottom rows indicate, respectively, the cluster distribution and the equivalent plastic strain field on the neutral z-plane.

We now compare the performance of the above two clustering methods with the online dynamic approach (see Section 3.2.2) which aims to refine clustering as an RVE is deformed and softening appears. The first step of the dynamic approach is to determine the number of new clusters per time step. To this end, we use our error-control adaptive scheme for new cluster generations relying on the AAF-IE's error metric in Equation (34), and we compare its performance against a naïve approach where the same number of new clusters are created per step. We set the initial and maximum numbers of clusters as 400 and 800, respectively. Both approaches use position-based clustering to create the initial 400 clusters but the additional 400 clusters are generated either by the error-controlled scheme or uniformly at each step.

In Figure 16(a), we illustrate the adaptive Newton step sizes on the left axis and find that they are reduced to their minimum at the fifth step when material properties transit from elasticity to plasticity. The steps gradually grow and plateau when most material points enter the plastic regimes. The number of new clusters added in each step is recorded on the right axis in Figure 16(a) and indicates that, while the simple approach adds 8 clusters in each step (a total of 400 additional clusters), the adaptive scheme adds all the new clusters during the first eight steps when phase transitions occur.

We compare the effective strain-stress responses between the two approaches in Figure 16(b) where it is clear that the adaptive scheme provides higher accuracy. The primary reason behind this result is that in history-dependent deformations the adaptive scheme reduces the errors that appear early in the simulations (due to insufficient discretizations) and are accumulated as the deformation progresses.

To further improve accuracy, we now use the fully adaptive strategy which leverages both the offline stress-informed clustering and the online dynamic clustering. As illustrated in Figure 16(b) the solution of the fully adaptive clustering is more accurate than the previous two cases (all approaches have the same total number of clusters) and matches the solution obtained via DNS.

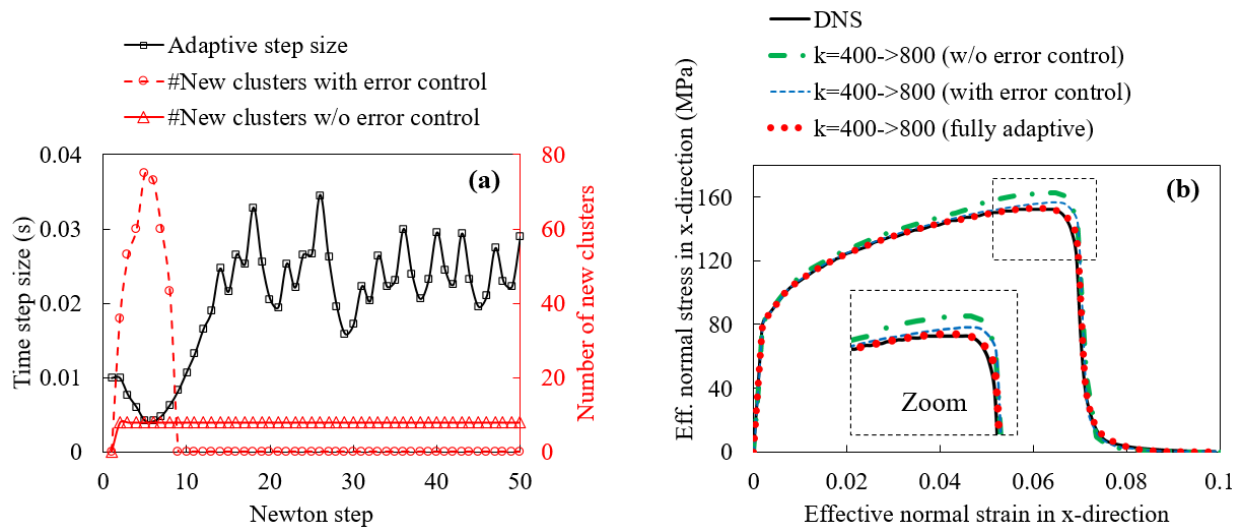

**Figure 16 Adaptive clustering with the error-control scheme: (a)** Adaptive Newton step sizes and the number of new clusters per step. **(b)** Accuracy improvements of the error-control adaptive schemes on homogenization results where the fully adaptive approach is the combination of offline stress-informed clustering with an online error-controlled adaptive counterpart.

Hereafter, for brevity, we refer to the clusterings without error control, with error control, and with full adaption as clustering type-1, 2, and 3, respectively. In the top row of Figure 17, we plot the spatial distributions of the cluster centroids across the three methods. We observe that the new clusters from type-2 and especially type-3 methods are quite compact and largely located near pore surfaces. This behavior is because when material properties change abruptly during phase transition, AAF-IE adaptively reduces time steps and generates more clusters at each step. The new clusters are created around pore surfaces where high solution gradients appear. As shown in the bottom row of Figure 17, this compact clustering behavior improves the accuracy in capturing concentrations.

We analyze the computational costs of type-3 (full adaption) clustering in Figure 18(a) where the offline and online stages consume 54.9 and 291.2 seconds, respectively. The online step time increases in early steps due to phase transition and then it drops as materials are yielded. The step time rises again in late steps due to damage evolutions where expensive microscopic damage models are activated to map effective responses between the reference microstructures used in damage modeling, see Appendix B.

Finally, we compare the online computational costs of different clustering approaches against the clock time of DNS in Figure 18(b). We note that as the clustering types advance from the position-based method to offline stress-informed and fully-adaptive approaches, the required number of clusters for matching DNS decreases. In addition, compared to DNS which relies on the pure implicit scheme with re-assembling stiffness matrix at each iteration, our adaptive ROM with temporal reductions (by AAF-IE) and spatial reductions (via adaptive clustering) needs less than 3% computational time.

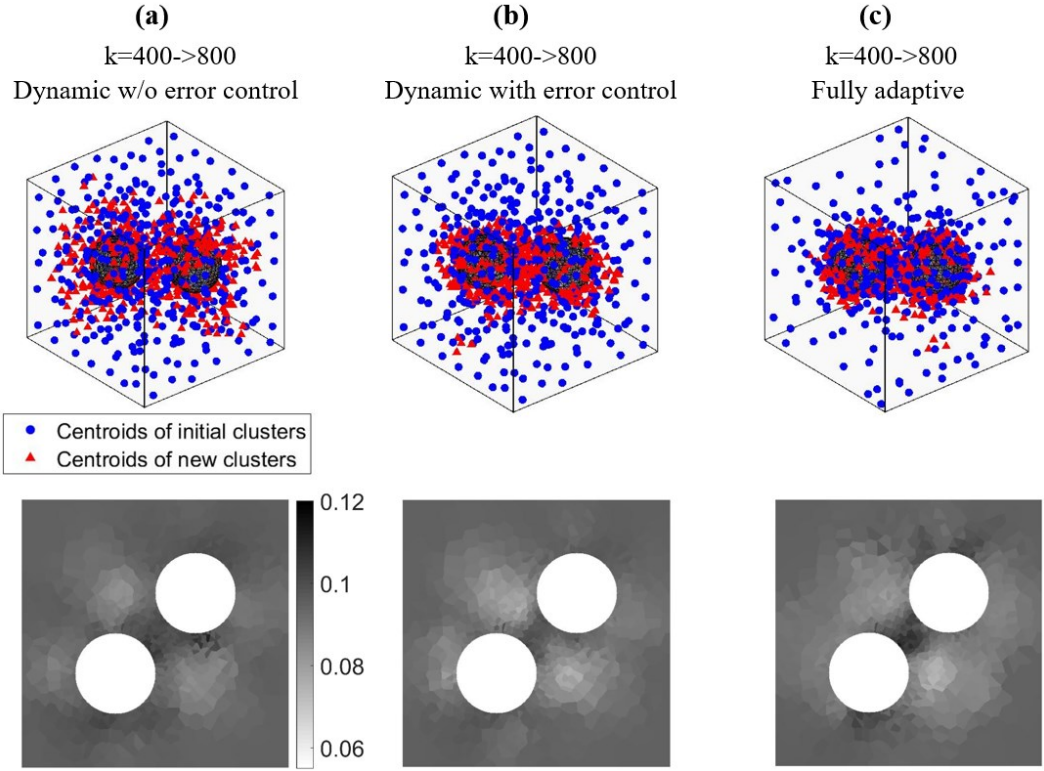

**Figure 17 Different clustering approaches and effective plastic strain distributions:** The top row indicates the spatial positions of the initial (offline) and dynamic (online) clusters across the three different clustering approaches. The plots in the bottom row illustrate the distribution of the effective plastic strains on the neutral z-plane for each of the approaches.

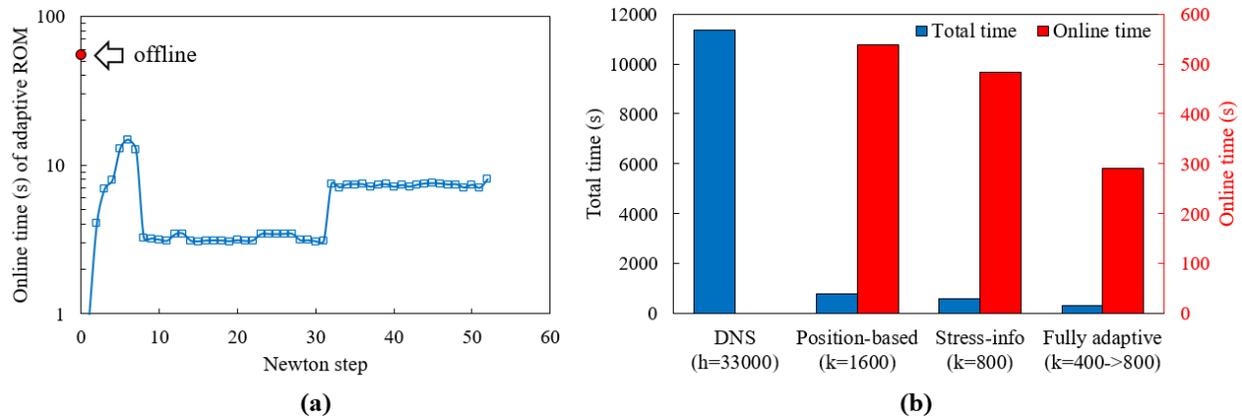

**Figure 18 Time analyses on the fully adaptive method:** **(a)** Online solution time per step of the fully adaptive ROM. **(b)** Time comparisons between DNS and ROMs (h: number of elements, k: number of clusters) where different clustering approaches converge to the same DNS solutions.

### 4.2.2. Effects of porosity volume fraction

Porosity volume fraction significantly affects alloys' failure behaviors [46]. To investigate its influence, we keep the positions and shapes of the two pores in the previous subsection identical but change their radii to 8 μm, 15 μm, 22 μm, and 30 μm. As shown in Figure 19(a), these four cases correspond to microstructures with porosity volume fractions of 0.43%, 2.8%, 8.9%, and 22.6%, respectively.

We deform these four RVEs using the same macro-deformation gradient in Equation (52) and illustrate their effective stress-strain responses in Figure 19(b). We observe that with the increase of porosity volume fractions, the values of both toughness and UTS are significantly reduced. Specifically, the toughness and UTS of the microstructure with 22.6% porosity are only 79.2% and 80.9% of the one with 0.43% porosity. In addition, by comparing our fully adaptive ROM against DNS, we find our ROM achieves high fidelity with solution errors smaller than 3%, as listed in Table 5.

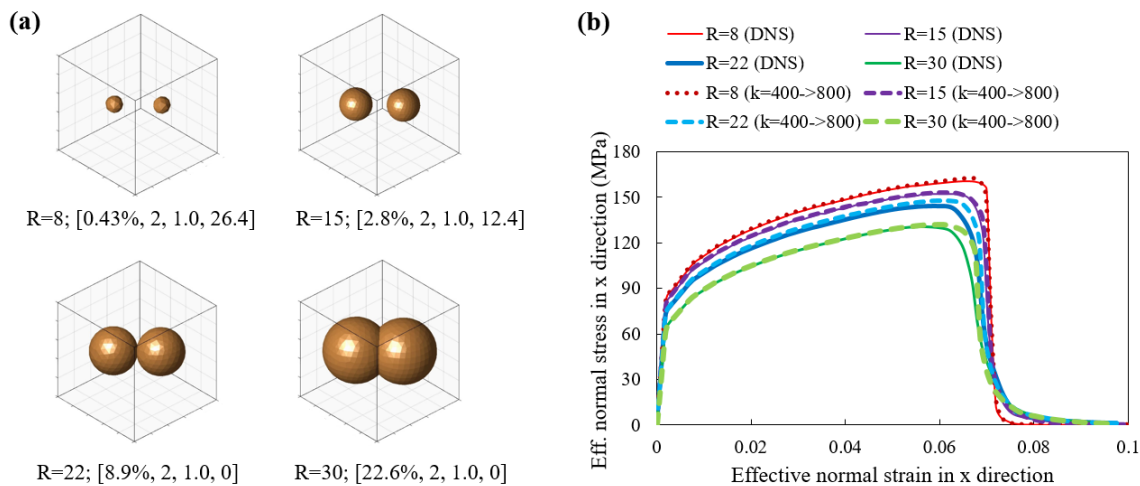

**Figure 19 Microstructures with different porosity volume fractions: (a)** Morphologies of the microstructures with distinct pore radii (R) and descriptors $[V_f, N_p, A_r, \bar{r}_d]$. **(b)** Homogenized stress and strain curves.

**Table 5**: **Accuracy analyses:** Errors of ROMs' toughness and UTS as compared to DNS.

| Pore radii (μm) | Error of toughness (%) | Error of UTS (%) |
| --- | --- | --- |
| 8 | 0.34 | 1.24 |
| 15 | 0.71 | 0.65 |
| 22 | 2.27 | 2.02 |
| 30 | 2.19 | 0.76 |

### 4.2.3. Microstructure with complex morphology

In this section, we test the performance of our ROM using the complex RVE in Figure 20(a) which is subject to the deformation gradient in Equation (52). The distributions of the equivalent plastic strain obtained by DNS and ROM with fully adaptive clustering are compared in Figure 20(b-e). In this simulation, the ROM is accelerated by both temporal and spatial adaption techniques. We see that ROM's solutions (strain distributions and localizations) converge to DNS as more clusters are used. However, compared to the RVE in Figure 11(a) which has relatively simple pore geometries, more clusters are needed in this test to capture local porosity morphology to ensure ROM's solution converges to that of DNS.

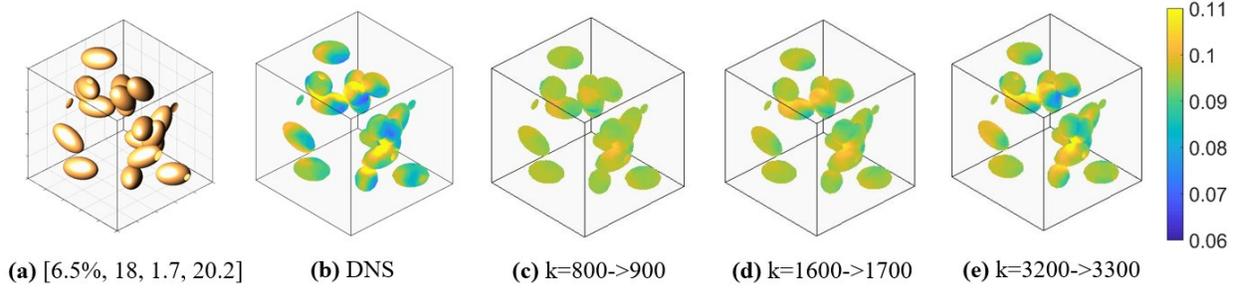

**(a)** [6.5%, 18, 1.7, 20.2]　　**(b)** DNS　　**(c)** k=800->900　　**(d)** k=1600->1700　　**(e)** k=3200->3300

**Figure 20 Microstructure with complex pore morphology: (a)** The studied microstructure and its pore descriptors $[V_f, N_p, A_r, \bar{r}_d]$. **(b)** Distribution of equivalent plastic strains emulated via DNS. **(c-e)** Distributions of equivalent plastic strains are simulated by adaptive ROMs with different numbers of clusters where the first and second numbers indicate the initial and total number of clusters, respectively.

　　The homogenized responses are compared between DNS and our ROM in Figure 21(a). When the initial number of clusters increases to 3200, the ROM's stress-strain curve matches DNS in terms of both UTS and fracture toughness. To further investigate the discrepancy between DNS and ROM, we plot an error histogram in Figure 21(b) which measures the difference in the effective plastic strains between DNS and ROM at each element. It is observed that compared to fewer clusters (e.g., k=200->300), the ROMs with more clusters have lower prediction errors. Predicted toughness values of DNS and ROM are compared in Table 6 where we see that the error drops from 1.61% to 0.56% as the total cluster numbers increase from 300 to 3300.

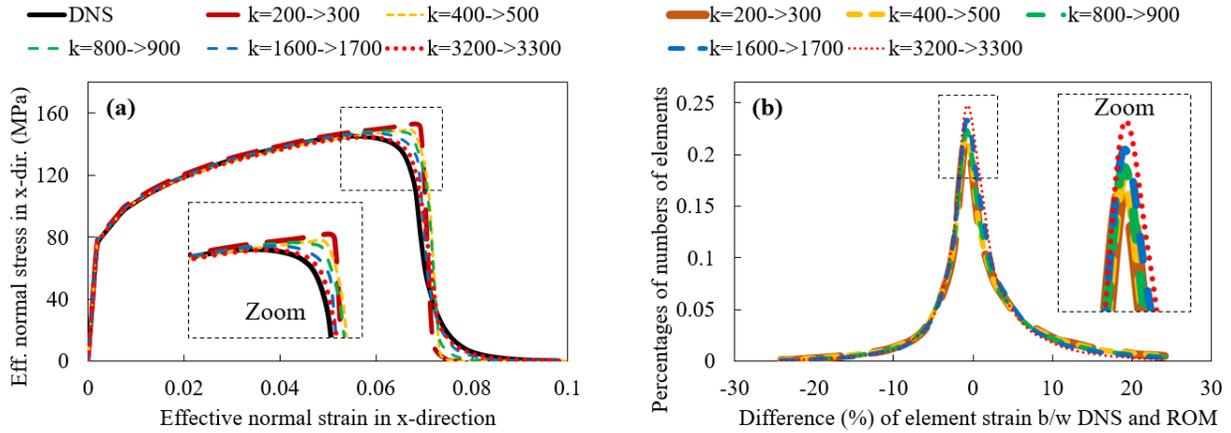

**Figure 21 Accuracy analyses on effective stresses: (a)** Comparisons of homogenization results between DNS and adaptive ROMs. **(b)** Element-to-element strain comparisons between DNS and ROMs.

**Table 6**: **Accuracy analyses on toughness:** Comparison of material toughness simulated by DNS and ROMs for the studied microstructure in Figure 20(a).

| Method | Toughness (MJ/mm³) | Error (%) |
|---|---|---|
| DNS | 8.89 | - |
| ROM (k=200->300) | 9.04 | 1.61% |
| ROM (k=400->500) | 9.01 | 1.29% |
| ROM (k=800->900) | 9.02 | 1.42% |
| ROM (k=1600->1700) | 8.95 | 0.61% |
| ROM (k=3200->3300) | 8.84 | 0.56% |

Finally, we compare the computational costs of DNS and ROMs in Figure 22. We note that the offline costs are excluded from the time analyses since they are only performed once and their results can be reused for any other deformation. We add that with the highest adaptive cluster numbers (k=3200->3300), the ROM's online time only accounts for about 1.3% of that of DNS; indicating an acceleration factor of 79.9 in this example.

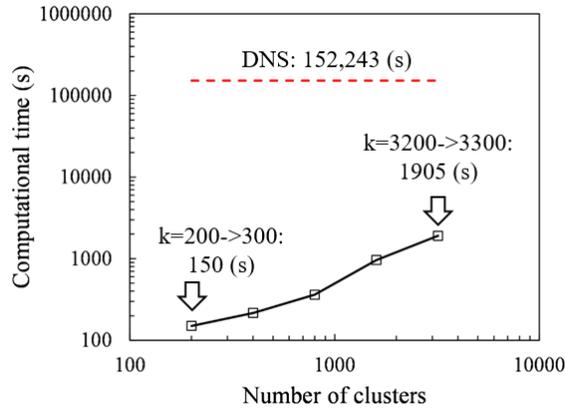

**Figure 22 Time comparisons between DNS and our fully adaptive ROMs:** Computational time is compared between DNS and ROMs with different numbers of adaptive clusters for the microstructure in Figure 20.

*4.3. Multiscale experiment*

In this section, we integrate the temporal and spatial adaption schemes with a FOCH-based concurrent multiscale framework to investigate the influence of micro-pores on the damage behavior of a macro component. Our multiscale study is performed on the same 3D plate model in Figure 6 which is now considered the macro-structure, see Figure 23(a). We assume the same Dirichlet boundary conditions are applied on the plate as in Figure 6(b). We also presume the plate consists of a monoscale region without any porosity and a multiscale region with microscopic pores, see Figure 23(b). To avoid softening-induced mesh sensitivity, we follow the mesh dependency study in Section 4.1 by discretizing the plate with 13000 linear tetrahedral elements and applying the non-local function with the fracture bandwidth of 1.5 mm. Under this discretization, the mono- and multiscale regions are meshed by 10560 and 2440 elements, respectively.

To model local morphologies, we associate all macro-IPs in the multiscale region with spatially varying porous microstructures. We highlight four IPs in the multiscale region whose corresponding microstructures are visualized in Figure 23(c). Although the four RVEs share the same porosity volume fraction of 6.5%, their local morphologies are significantly different and vary from having one spherical pore to having multiple randomly dispersed overlapping pores. These four types of RVEs are randomly assigned to the macro-IPs. To ensure consistent fracture energies across scales [19], we enforce the sizes of RVEs to be the same as the sizes of macroscale elements.

We mesh the microstructures of our multiscale model based on the complexity of their morphologies. Specifically, we mesh the four microstructures associated with macro-points A, B, C, and D with 15000, 78000, 103300, and 60400 elements, respectively. Correspondingly, our multiscale model is discretized with a total of roughly 156 million elements. Since the computational cost of simulating such a model via DNS (FE$^2$) is prohibitively high, we only utilize our ROM where we assign 300 adaptive clusters to the microstructure with a single spherical pore and 900 clusters to all other microstructures.

The computational time of our concurrent multiscale simulation is about 59.4 hours. Based on the time comparison between DNS and ROMs in Sections 4.1 and 4.2, the estimated time for DNS (FE$^2$) is more than 52519.9 hours (2188.3 days) given the same computational resources. That is, our method achieves an acceleration factor of 884.2 in this multiscale study.

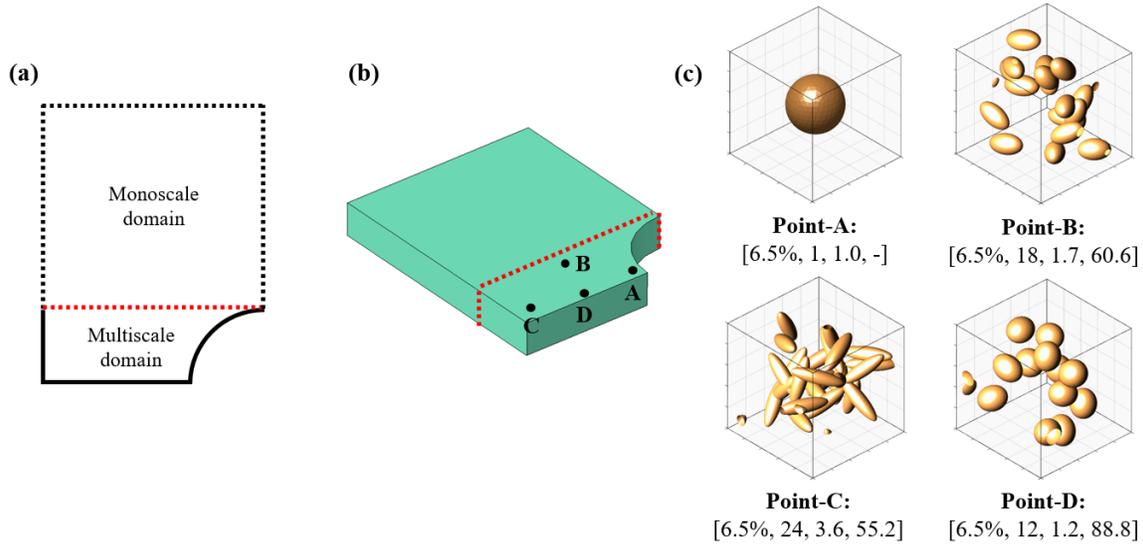

**Figure 23 Multiscale model: (a)** The macroscale component comprises a monoscale region without any porosity and a multiscale region containing micro-pores. **(b)** All integration points are assigned with porous microstructures in the multiscale region with four highlighted points. **(c)** The four points are assigned with different microstructures, respectively, where their pore descriptors are listed in the vectors $[V_f, N_p, A_r, \bar{r}_d]$.

We demonstrate the multiscale model's load-displacement response in Figure 24(a) where the plate's ultimate load carrying capacity is 3659.1 N at the displacement of 0.24 mm, which is an 8.5% decline from its dense counterpart in Figure 9 where the plate ruptured at the displacement of 0.34 mm. We observe a localized fracture band that lies in the multiscale region and is oriented orthogonal to the Dirichlet boundaries in Figure 24(b). Two of the highlighted material points (B and C) are close to the fracture band while the other two points (A and D) are distant from the softening regions. The microscale equivalent plastic strain fields in the corresponding four RVEs are demonstrated in Figure 24(c) where we observe that microstructures B and C have accumulated much larger strains (note the axis scales). In each of these RVEs, we notice that high plastic strains concentrate in regions where pores are closely packed.

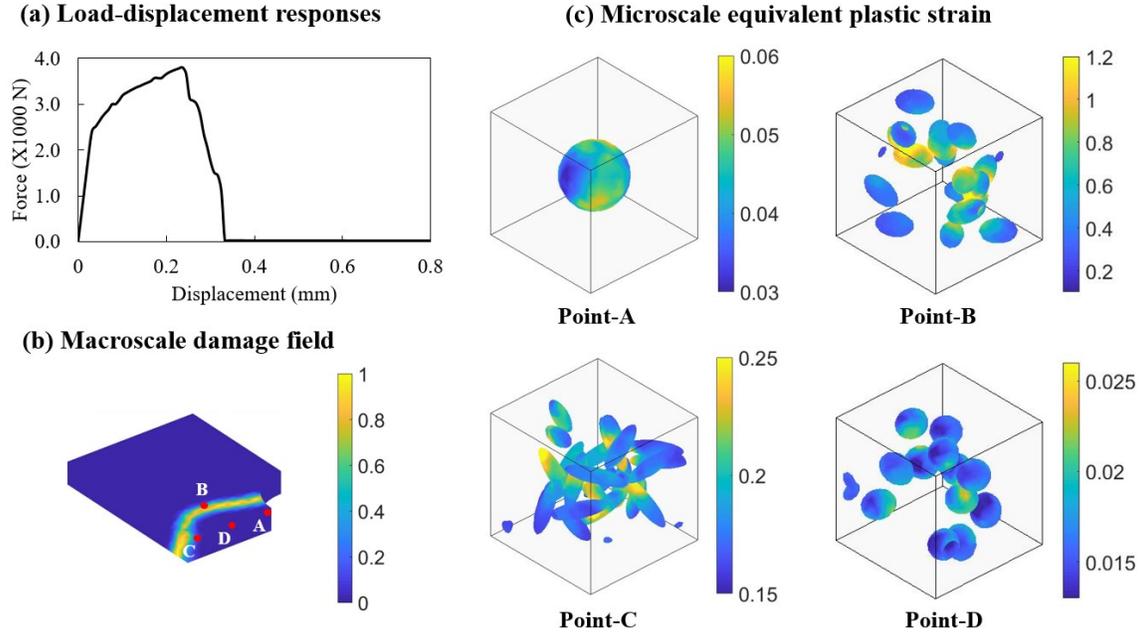

**Figure 24 Results of the adaptive ROM: (a)** Macroscale load-displacement responses. **(b)** Macroscale damage patterns with a fully developed fracture band connecting mid-hole to outer surfaces. **(c)** Microscale distributions of the equivalent plastic strains are associated with the four integration points on the macroscale model.

## 5. Conclusion

We introduce adaptive deflated clustering analysis in this paper which is a novel ROM that can simulate the damage behavior of metallic alloys with process-induced spatially varying microscopic pores. Our method is comprised of two new adaption strategies that are readily extensible for any clustering-based ROMs: a novel AAF-IE temporal adaption scheme and a new spatial clustering approach. Our AAF-IE not only alleviates softening materials' instabilities by preserving the positive definiteness of the underlying algebraic system's stiffness matrices, but also improves the overall efficiency by avoiding stiffness matrices' re-assembly during the online process. In addition, our adaptive spatial discretization approach significantly improves the ROM's prediction accuracy over critical regions by selectively modifying local interpolations. Specifically, the new clustering approach automatically adjusts cluster densities according to online solutions and resizes temporal steps to allocate more computational resources at crucial time instances. In our ROM we use a macroscale non-local function and a microscopic energy-based damage formulation to resolve softening-induced pathological mesh dependencies and ensure fracture energy consistency across scales.

We test the performance of our adaptive ROM via several numerical experiments. Our ROM results show significant improvements in robustness, efficiency, and local accuracy compared to DNS. We also apply the proposed method to a concurrent multiscale model to simulate the multiscale softening responses and indicate that components with spatially varying porosity defects have lower ultimate load-carrying capacity than their dense counterparts.

We note that There are some fundamental differences between our ROM with existing techniques. As an example, contrary to SCA which groups elements based on their mechanical responses, our ROM agglomerates elements based on their geometrical proximity. Another major difference is our novel hybrid integration scheme that avoids softening-induced solution divergences. Additionally, SCA is typically applied to composite or polymer materials [14], [15] with strong or weak inclusions where the property ratio between material phases (e.g., moduli) is reasonably small. Our ROM is used to simulate alloys with pores where the moduli difference between material and void is infinite. Compared with strong or weak inclusions, it is much harder to simulate microstructures with pores (especially for an FFT-based approach whose computational efficiency significantly deteriorates due to the infinite property contrast between material phases).

Given the promising results from this work, our ROM can be extended to finite strains and extreme load conditions in the future. Another research direction of interest is to use our ROM as an efficient database generator for complex material responses, followed by deep learning approaches [47]–[50] to directly correlate materials' local morphologies with their responses. It is also noted that in this work the plasticity behavior of the metallic matrix is based on the J2 plasticity model as opposed to, e.g., crystal plasticity [51]–[53]. Adding crystal plasticity to our multiscale model is an interesting idea which we plan to explore in our future works.

## Acknowledgments

The authors thank the ACRC consortium members. Specifically, we appreciate Randy Beals from Magna International, and Chen Dai from VJ Technologies for providing us with the W-profile plate samples and X-ray computed tomography scanning and data generation, respectively. Ramin Bostanabad also acknowledges NSF funding (award numbers OAC-2211908 and OAC-2103708).

## Appendix A. Deflated clustering analysis

A computationally efficient reduced-order multiscale model, named deflated clustering analysis (DCA), is proposed in [17] to simulate the elastoplastic behaviors of metallic alloys with process-induced, spatially varying porosity defects. DCA relies on the clustering-based domain decomposition that universally applies to multiscale domains to accelerate the accurate computations of macroscale deformations and effective microscopic responses.

Compared to the classic $FE^2$ approach, which solves equilibrium equations on both macro- and micro-scales via computationally expensive FEM, the numerical advantage of DCA is to accelerate emulations by systematically agglomerating neighboring materials points to clusters and projecting solution variables into clustering-based deflation spaces for efficient nonlinear simulations, see Figure A.1. While both macro- and micro- computations follow general FEM formulation with displacements as unknown variables, the acceleration mechanisms on macro- and micro-domains are different.

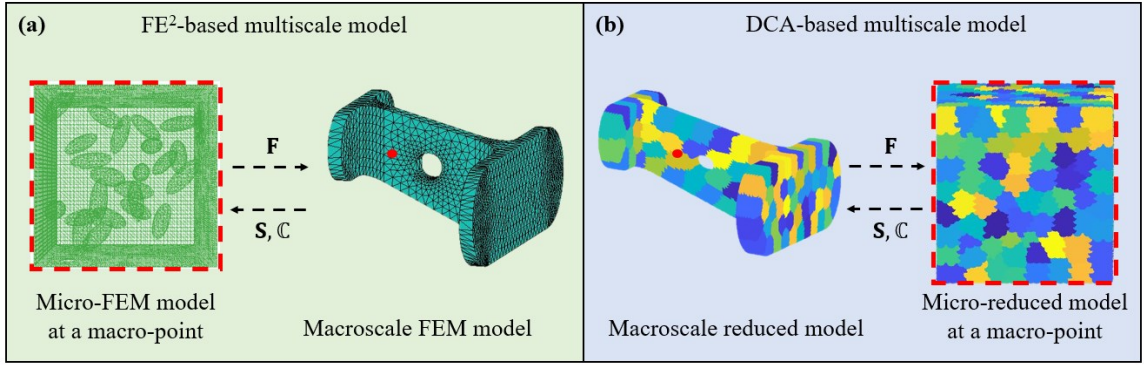

**Figure A.1 Comparison of the classic multiscale method with DCA: (a)** Classic FE$^2$-based multiscale model assumes each macroscale material point is associated with a microstructure where the deformation gradients (**F**) are passed from the macro- to micro-domains, and the effective stress (**S**) and moduli ($\mathbb{C}$) are passed backward. **(b)** DCA speeds up nonlinear simulations on multiple scales via clustering-based data compression to systematically reduce unknown variables where different clusters are marked by distinct colors.

Macroscale clustering is integrated with the deflation method to accelerate the conjugate gradient (CG) process of the underlying algebraic systems at each Newton iteration. The deflation method aims to remove the near-zero eigenvalues from the algebraic system's stiffness matrices, and it simultaneously lowers their conditional number which, in turn, increases the computational efficiency substantially. An incremental deflation method is also utilized to avoid any unnecessary re-assembly of the deflation matrix during runtime. FEM solutions are recovered from the deflation space per CG iteration, and their convergence errors are checked based on prescribed tolerance. By enforcing the same CG convergence criteria between FEM and deflation method, the macroscale ROM converges to the same displacement results as those of FEM without losing solution accuracy. Therefore, the local deformation gradients at each macroscale integration point are highly accurate for the sequent microscale analyses.

For microscale, clustering aims to generate coarse computational grids, similar to the coarse-graining process, where node-to-node interactions of FEM are condensed to the cluster interactions in the reduced cluster-based mesh. For the microscale equilibrium equations, utilization of more clusters results in a larger algebraic system with more eigenmodes and sophisticated deformations but with the expense of higher degrees of freedom and longer simulation time. In essence, the microscale clustering analysis targets to provide homogenized responses at macro-integration points by coarse-graining responses from micro-clusters. The efficiency of the microscopic ROM is much improved than FEM since degrees of freedom are dramatically reduced from large numbers of elements to a few clusters. Although it loses local accuracy as the solution variables are assumed uniform within each cluster, it is sufficient for the sake of homogenizations.

Detailed steps of applying DCA in microscale simulations are summarized in Algorithm A.1. We note that following the construction of a clustering-based reduced mesh and stiffness matrix, microstructural local stress and strain fields are averaged at the centroid of each cluster. In other words, instead of computing distinct solution fields at different elements, close-by material points (in one cluster) are assumed to share the same solutions. We also stress that DCA in [17] is dependent on naïve position-based clustering, and its accuracy on localized solutions can be improved by the proposed adaptive clustering technique in this work.

**Algorithm A.1** Structure of the microscale cluster-based ROM

```
 1: procedure Solve for the homogenized microstructural responses at the i^th nonlinear increment
 2:     ▷Initialization
 3:     if i ← 1, then
 4:         Generate FE mesh on the microstructure
 5:         Compute the microstructure geometric center x_center from nodal coordinates x
 6:         Create node-based clusters via domain decomposition
 7:         Construct cluster-based reduced mesh
 8:         Initialize material properties with elastic material properties
 9:         Develop reduced stiffness matrix on cluster-based mesh
10:     else
11:         Update material properties from the last increment
12:         Update the cluster-based constitutive model
13:     end if
14:     Read macroscopic deformation gradient F^i(X)
15:     Compute incremental homogeneous displacements on FE mesh: u_0^i(x) = (F^i(X) − I) · (x − x_0)
16:     Project the homogeneous displacements from FE mesh to cluster-based mesh
17:     ▷Starting Newton iterations to solve micro-equilibrium equations on the reduced mesh
18:     Set Newton iteration convergence criteria ϵ = 10^−3
19:     for j ← 0, N do       ▷Loop over N Newton iterations
20:         Compute the microscale internal force vector f_j^int(u_j^i(x))
21:         Solve micro-equilibrium for microscopic displacement fluctuation ũ_j(x)
22:         Update microscale displacements: u_{j+1}^i(x) = u_j^i(x) + ũ_j(x)
23:         Postprocess for microstructural stress and strain fields
24:         ▷check for convergence
25:         if iteration residual < ϵ , then
26:             Compute homogenized stress tensor and material modulus
28:             return
29:         end if
30:     end for
40: end procedure
```

## Appendix B. Stabilized micro-damage model

Strain softening causes severe convergence difficulty to implicit integration procedures, e.g., the Newton-Raphson method, see Section 2.4. Although our impl-exp integration scheme can successfully resolve softening-induced instability, the global stiffness matrix of the underlying algebraic system still needs to update the entries corresponding to damaged elements amid crack propagations. And as large numbers of elements enter damage realms, the updates are expensive. To further improve the microscale computational efficiency, we adopt the stabilized micro-damage model [31] which is reviewed in this section.

We first demonstrate the classic progressive damage model in Figure B.1(a) where the damage manifests itself in both material moduli degradation and yield stress reduction. Specifically, as an arbitrary material point is subject to strain softening, its damaged yield stress ($\mathbf{S}_m^d$) and degraded modulus ($\mathbb{C}_m^d$) are computed by projecting its undamaged reference stress ($\mathbf{S}_m^{ref}$) and the intact elastic modulus ($\mathbb{C}^{el}$) by damage variables, see Section 2.2. We point out that this model focuses on the damage evolutions on a single integration point, and the subscript 'm' presents an arbitrary material point in a microstructure.

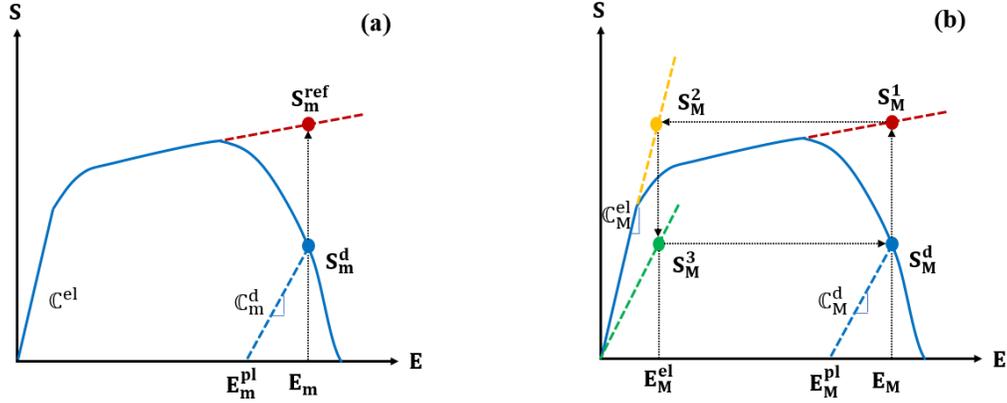

**Figure B.1 Comparison of damage models: (a)** Classic progressive damage model at each integration point of RVE. **(b)** Stabilized micro-damage model computes the homogenized damage responses of the original RVE by introducing three additional reference RVEs that partially share their state variables.

In contrast, the stabilized micro-damage model [31] aims to emulate an RVE's softening by considering its overall response. Its advantage is that it projects the damage evolutions from plastic hardening by introducing three additional reference RVEs with partially shared state variables, and the whole procedure is simply based on pure implicit schemes without convergence issues. To further improve this method's efficiency, we replace its implicit scheme with our AAF-IE for computing RVE's effective damage responses.

For an arbitrary RVE subject to strain softening, its homogenized stress-strain response can be demonstrated by the blue curve in Figure B.1(b) where its damaged homogenized stress $\mathbf{S}_M^d$ is computed as:

$$\mathbf{S}_M^d = \mathbb{C}_M^d : \mathbf{E}_M^{el} = \mathbb{C}_M^d : (\mathbf{E}_M - \mathbf{E}_M^{pl}) \tag{B.1}$$

where $\mathbb{C}_M^d$ is the homogenized elastic modulus of the damaged RVE, $\mathbf{E}_M$, $\mathbf{E}_M^{el}$ and $\mathbf{E}_M^{pl}$ are the RVE's effective total strain, elastic strain, and plastic strain, respectively, where the subscript 'M' indicates the associated variable is homogenized for overall macroscale property. The un-damaged stress of the first reference RVE can be computed by Equation (B.2), see the red curve in Figure B.1(b), where we assume the referenced RVE shares the same elastoplastic property as its original counterpart but in absence of damage.

$$\mathbf{S}_M^1 = \mathbb{C}_M^{el} : \mathbf{E}_M^{el} = \mathbb{C}_M^{el} : (\mathbf{E}_M - \mathbf{E}_M^{pl}) \tag{B.2}$$

where $\mathbb{C}_M^{el}$ is the shared effective elastic moduli of the original and the first reference RVEs. Combining Equation (B.1) and (B.2), the stress states on the two RVEs are related by:

$$\mathbf{S}_M^1 = \mathbb{C}_M^{el} : (\mathbb{C}_M^d)^{-1} : \mathbf{S}_M^d \tag{B.3}$$

Let homogenized stress of the first RVE ($\mathbf{S}_M^1$) identical to that of the second RVE ($\mathbf{S}_M^2$) which we assume to have the same material properties as the original RVE but it only deforms elastically, see the yellow curve in Figure B.1(b). The elastic macroscale strain of the second RVE is then computed by inverting the elastic modulus of Equation (B.2) as:

$$\mathbf{E}_M^{el} = \left(\mathbb{C}_M^{el}\right)^{-1} : \mathbf{S}_M^1 = \left(\mathbb{C}_M^{el}\right)^{-1} : \mathbf{S}_M^2 \tag{B.4}$$

where we assume microscale stress of the second RVE to satisfy elastic constitutive equation as:

$$\mathbf{S}_m^2 = \mathbb{C}^{el} : \mathbf{E}_{m2}^{el} \tag{B.5}$$

where $\mathbf{E}_{m2}^{el}$ represents the local elastic strain at an arbitrary integration point in the second referenced RVE, and the homogenized elastic strain can be computed via the volume averaging process as:

$$\mathbf{E}_M^{el} = \frac{1}{|\Omega|} \int_\Omega \mathbf{E}_{m2}^{el} d\Omega \tag{B.6}$$

where $\Omega$ and $|\Omega|$ are the domain and volume of the original RVE which are shared among all reference RVEs. We assume the local value of the elastic strain ($\mathbf{E}_{m2}^{el}$) is shared between the second RVE and the third RVE, which only deforms elastically but its effective elastic modulus is identical to that of the damaged original RVE ($\mathbb{C}_M^d$), see the green curve in Figure B.1(b). Under this assumption, the local stress of the third RVE ($\mathbf{S}_m^3$) is computed by:

$$\mathbf{S}_m^3 = (1-D_m)\mathbb{C}^{el} : \mathbf{E}_{m2}^{el} = (1-D_m)\mathbb{C}^{el} : \mathbf{E}_{m3}^{el} \tag{B.7}$$

where $D_m$ is the local damage parameter at a microscale integration point. Its value is determined by the state variable from the first reference RVE. Towards the end, the elastic microscale stress in the third RVE is homogenized, which equals the effective stress of the original damaged RVE as:

$$\mathbf{S}_M^d = \frac{1}{|\Omega|} \int_\Omega \mathbf{S}_m^3 d\Omega \tag{B.8}$$

In addition, an effective damage parameter can be defined by the homogenized stresses from the original and the first RVEs as:

$$D_M = 1 - \frac{\|\mathbf{S}_M^d : \mathbf{S}_M^1\|}{\|\mathbf{S}_M^1 : \mathbf{S}_M^1\|} \tag{B.9}$$

where $D_M$ is the homogenized damage parameter that indicates the damage status of the macroscopic integration point associated with the studied RVE. Similar to the single-scale damage parameter defined in Section 2.2, $D_M$ ranges between [0,1]. The macro-point is considered fully ruptured when $D_M = 1$, even if only parts of the elements in the associated RVE are damaged.